\documentclass{article}
\usepackage[utf8]{inputenc}
\usepackage{commath}
\usepackage{array}
\usepackage{tabu}
\usepackage{graphicx}
\usepackage{subfig}
\usepackage{xcolor}
\usepackage{float}
\usepackage{amsthm}
\usepackage{amsmath}
\usepackage{amssymb}
\newtheorem{theorem}{Theorem}[section]

\usepackage{hyperref}
\usepackage{epstopdf}

\title{Bioeconomic analysis of harvesting within a predator-prey system: A case study in the Chesapeake Bay fisheries}
\date{December 2022}

\author{I.N. Panayotova, J. Herrmann, and N. Kolling}

\begin{document}
\maketitle

\begin{abstract}
Sustainable use of biological resources is very important as over exploitation on the long run may lead to stock depletion, which in turn may threaten biodiversity. The Chesapeake Bay is an extremely complex ecosystem, and sustainable harvesting of its fisheries is essential both for the ecosystem's biodiversity and economic prosperity of the area. Here, we use ecosystem based mathematical modeling to study the population dynamics with harvesting of two key fishes in the Chesapeake Bay, the Atlantic Menhaden (\textit{Brevoortia tyrannus}) as a prey and the Striped Bass (\textit{Morone saxatilis}) as a predator. We start by fitting the generalized Lotka-Volterra model to actual time series abundance data of the two species obtained from fisheries in the Bay.  We derive conditions for the existence of the bio-economic equilibrium and investigate the stability and the resilience of the biological system. We study the maximum sustainable yield, maximum economic yield, and resilience maximizing yield policies and their effects on the fisheries long term sustainability, particularly with respect to the menhaden-bass population dynamics. This study may be used by policy-makers to balance the economic and ecological harvesting goals while managing the populations of Atlantic menhaden and striped bass in the Chesapeake Bay fisheries.
\end{abstract}


\section{Introduction}\label{sec1}

The Atlantic menhaden (\textit{Brevoortia tyrannus}) is widely considered to be one of the most important species of fish in the Chesapeake Bay. Atlantic menhaden grow up to 18 inches, can live up to around 10 to 12 years, and can weigh up to a pound \cite{popBioMenhaden, cbp2}. They generally move around in schools, and their spawning will generally take place during cooler months but \emph{can} occur year round \cite{popBioMenhaden, popFishMenhaden}. The Atlantic menhaden is a filter-feeder, and its diet consists mostly of various types of plankton and plant materials \cite{cbp2, MenJuvFood, noaa}. The Atlantic menhaden is an extremely important fish in the Chesapeake Bay, providing both biological and economic benefits. It serves to transfer nutrients from lower to higher trophic levels in various food chains in the Bay and is a key prey species for other fish, such as the striped bass \cite{bassEatMen}, bluefish \cite{blueEatMen}, weakfish \cite{weakEatMen}, and others \cite{cbp2, noaa}. Along with this, the oil produced from the Atlantic menhaden has both economic and health benefits. It is a good source of Omega-3 fatty acids, which may help with cardiovascular diseases \cite{noaa}.

However, the overall population of the Atlantic menhaden has been on a significant decline since the 1970s and continues to remain low \cite{cbf}. Species within the Chesapeake Bay, particularly Atlantic menhaden, are negatively affected by environmental issues such as air pollution, invasive species in the ecosystem, wetland destruction from development, deforestation, climate change, rising sea levels, runoff pollution, and chemical contaminants \cite{CBFissues, CBPissues, NPSissues, NWFissues}. However, one of the largest threats to the Atlantic menhaden is over harvesting \cite{cbf}. Large corporations have been in the spotlight for many years for ignoring harvest caps placed on the Atlantic menhaden \cite{ cbf, DPressMenhaden, TRCP, VPMenhaden}. Even with all these environmental and commercial factors that contribute to increasing the risk of further depletion of Atlantic menhaden, the Atlantic States Marine Fisheries Commission stated in their 2019 Stock Assessment Report that ``the Atlantic menhaden is not overfished and overfishing is not occurring" \cite{MenhadenStock}.

Another critical fish within the Chesapeake Bay is the striped bass (\textit{Morone saxatilis}).  The adults can range anywhere from 2 to 3 feet in length and 10 to 30 lbs in weight \cite{cbp, govSB}. The striped bass feeds on various prey, such as the Atlantic menhaden, crustaceans, and anchovies \cite{cbp, foodSB, bassEatMen}. They are also prey for other species in the Chesapeake Bay, such as sharks \cite{sharkDiet}, ospreys \cite{ospreyDiet}, and other larger species of fish \cite{cbp}. Striped bass spawning will generally take place from April through June, where the females will typically lay their eggs in fresh waters. From there, the juvenile striped bass will spend a few years around the areas they hatched in \cite{cbp, sbSpawns,govSB}. The most important roles of the striped bass in the Bay are its role in the fisheries food chain, because it is prey to larger predators and a key predator to smaller species, as well as its role in supporting one of the Bay's largest commercial and recreational fisheries \cite{cbp}. A study conducted by the Southwick Associates for the McGraw Center for Conservation Leadership analyzed the economic contributions of both the recreational and commercial fisheries for the striped bass \cite{mcgraw}. Striped bass recreational fisheries support a large portion of the gross domestic product (GDP), number of jobs supported, and majority of commercial landings in fisheries for eleven different states along the Atlantic coast \cite{mcgraw}. 

Despite the importance of the striped bass, striped bass populations have been at risk of extinction many times in the past several decades. The striped bass population faced a significant decline throughout the 1970s and 1980s. In addition to environmental factors such as pollution, acidity in bass spawning areas, and low dissolved oxygen in the bass’s habitats, overharvesting is the largest factor contributing to the striped bass population's decline \cite{sbli}. Other issues that negatively affect striped bass populations include population decline in the striped bass' prey, such as the Atlantic menhaden;  climate change; and even poor recreational handling of the caught fish - it is estimated that up to 9\% of released bass dies \cite{sbrf}. According to the 66th Northeast Regional Stock Assessment Workshop Assessment Report by the Northeast Fisheries Science Center, the striped bass stock was overfished and experienced overfishing in 2017 \cite{SBsa}.

Renewable resources and ``common property" fisheries are frequently exploited at levels that do not account for their effect on the ecosystem and long term sustainability, which may lead to extinction of some or all overexploited species. Reducing overexploitation and ensuring fisheries conservation for future generations is a main challenge for policy-makers, and mathematical models could help in understanding the complexities involved in this endeavor. 

Here we use the generalized Lotka-Volterra predator-prey model to study the population dynamics of Atlantic menhaden and striped bass and the possibility of balancing economic gains with ecosystem conservation. Even though this model is quite simple and is a great simplification of such a complex ecosystem such as the Chesapeake Bay, it is a useful tool that could provide some insights on the long-term dynamics of these two very important species. This simple model is analytically tractable and allows us to perform a comprehensive theoretical analysis of the most widely used harvesting policies within the predator-prey model. Taking into account that many fisheries are still managed as a single species and do not consider the effect of predator-prey interactions, we hope to deepen the understanding of the relationships between harvesting and sustainability within this simple predator-prey system with our analysis.

We use actual time series population data of the two fish to fit the mathematical model and estimate the model's parameters. We also took into account the current harvesting of the two species when fitting the model. The obtained model is then used to study the effects of harvesting on the economic rent function and long term sustainability of the fish population. We find the bioeconomic equilibria of the model as well as the stability of the equilibria. Along with this, we analyze the effects of applying the maximum sustainable yield (MSY) policy and the maximum economic yield (MEY) policy on our system, and we find how these strategies might be useful to both maximize profit from harvesting and keep a sustainable fish population. We also study the effects of resilience maximizing yield (RMY) policy when applied to the system. This policy is a management approach that prioritizes resilience, a measure of how fast the system stabilizes back to biological equilibrium after disruption, and allows fisheries to preserve the longevity and stability of the fish population over a long time. Finally, we perform sensitivity analysis to measure how small perturbations of our parameter values affect the population dynamics of the two species in time. This analysis allows us to identify the most sensitive parameters of the system, which could be then used for management and conservation decisions by policy-makers.

We must also mention that the model complexity could be increased in further studies to include more species or more environmental factors. However, increasing the complexity of the model is likely to change the quantitative and possibly the qualitative behavior of the system, including the responses of these two fishes to harvesting efforts. Nevertheless, this study is a useful contribution as it analyzes the most frequently used management policies in fisheries and accounts for the predator-prey interactions between Atlantic menhaden and striped bass. It also connects the theoretical investigations to actual time series data of the two species, and hence, it may be directly used by the decision makers to guide them for future management strategies.        

\section{Mathematical Model}\label{sec2}

\subsection{Two-species predator-prey model}
We consider the generalized Lotka-Volterra model describing the population dynamics of a prey $x=x(t)$ and a predator $y=y(t)$ given by the following system of differential equations:

\begin{equation}\label{LV_model}
\left\{
\def\arraystretch{2}
\begin{array}{lll}
\dfrac{dx}{dt} &=& x(r-ax-by) \\
\dfrac{dy}{dt} &=& y(-e-cy+dx).
\end{array}
\right.
\end{equation}

Here, $r$ is the intrinsic growth rate of the prey, $e$ is the death rate of the predator in absence of prey, $a$ and $c$ are the self-limitation parameters for the prey and the predator respectively, $b$ is the predator effect of $y$ on $x$, and $d$ is the effect of prey consumption of $x$ on $y$.
 This basic predator-prey model is a large simplification of the complex interactions between the Atlantic menhaden and striped bass in the Chesapeake Bay, under the following assumptions:
  \begin{enumerate}
\item In the absence of predator, the Atlantic menhaden grows logistically.  
\item The striped bass depends only on Atlantic menhaden for food.
\item In the absence of prey, the striped bass decreases logistically.
\item No interactions with other species or environmental factors are considered.
\item No age or size of the fishes is taken into account.
\end{enumerate}
Despite the simplicity of the Lotka-Volterra model, it is a primary model that captures the well known phase-shifted behavior of predator-prey interactions, it is analytically tractable, and it allows us to estimate the model's parameters with actual data. In this study, we aim to use this relatively simple model to gain insight into the effects of predator-prey interactions on the most frequently used harvesting policy strategies in fisheries management and hence, gain deeper understanding of the way multispecies food webs respond to harvesting of species at different trophic levels. 

\subsection{Data and Methodology}
The data used was the population abundance estimates in the Atlantic Menhaden 2019 Benchmark Stock Assessment and the 2022 Atlantic Striped Bass Stock Assessment Update Report from the years 1995 to 2017 \cite{ASMFC}. We adjusted the data to include harvesting by adding the estimates of commercial and recreational fishing efforts for the two species. We used MATLAB and Python to fit the model (\ref{LV_model}) to the time series data via the nonlinear least squares method. Due to the nonlinear nature of the fitting problem and being an over-determined mathematical system, there may be different sets of parameters providing similarly good fit to the actual time series data. Setting the upper bound for all parameters to 10 gave us the best fit for $r$, the intrinsic growth rate of Atlantic menhaden. The obtained parameter $r=0.513$ was of a magnitude similar to the intrinsic growth rate for Atlantic menhaden estimated from observational data using statistical analysis in \cite{balancing_model_complexity_ERP}.

\subsection{Results and Prediction analysis}
\begin{figure}[!ht]
    \centering
    \includegraphics[scale=0.65]{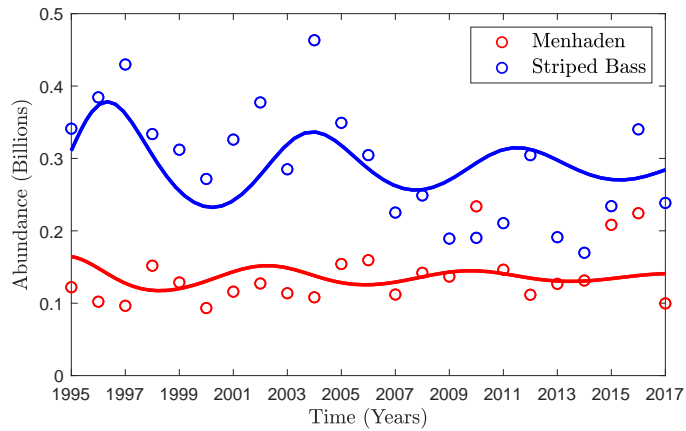}
    \caption{Fitted curves to the given data for the population of menhaden and bass from 1995 to 2017.}
    \label{Figure_Fitted_Model}
\end{figure}

To perform the fitting of the Lotka-Volterra model to the time series data for Atlantic menhaden and striped bass, we scaled the data down by $10^9$ for easier computation. The fitted model is graphed in Figure \ref{Figure_Fitted_Model} by the red and blue curves for the Atlantic menhaden and striped bass respectively. We can see the phase-shifted behavior of the Atlantic menhaden and striped bass, which portray the natural fluctuations commonly seen within predator-prey dynamics. The system's parameters obtained after the fitting are given in Table \ref{ParmsTable}, with a residual standard error $S = 0.11$. $S$ represents the average distance that the observed values fall from the regression line or how precise the regression model is on average, using the units of the response variable. As $S=0.11$, this tells us that the average distance of the data points from the fitted line is about 0.11.

\begin{table}[h!]
\centering
\begin{tabular}{|l|l|l|} 
\hline
Parameter & Description & Value\\
\hline
r & Intrinsic growth rate of Atlantic menhaden & 0.513\\
\hline
a & Self-limitation of Atlantic menhaden & 0.026\\
\hline
b & Effect of Striped bass on Atlantic menhaden & 1.765\\
\hline
e & Death rate of striped bass & 1.213\\
\hline
c & Self-limitation of striped bass & 0.520\\
\hline
d & Effect of Atlantic menhaden on striped bass & 9.999\\
\hline
\end{tabular}
\caption{Table of the model's parameter values obtained from the fitting.}
\label{ParmsTable}
\end{table}

\begin{figure}[!ht]
    \centering
    \includegraphics[scale=0.65]{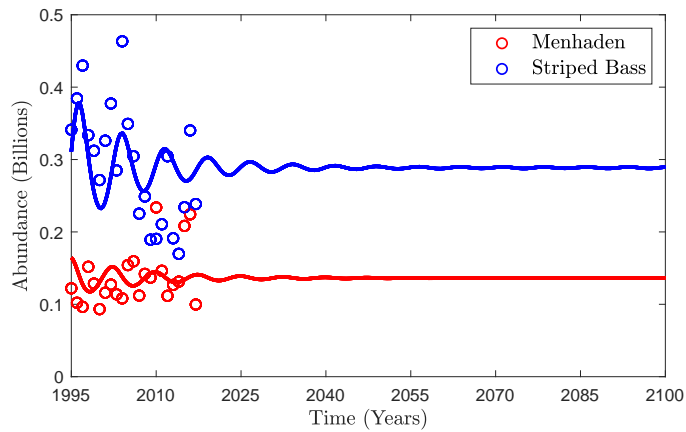}
    \caption{Model prediction of menhaden-bass population dynamics up to 2100.}
    \label{predictGraph}
\end{figure}

Next, we used the model and the obtained parameters to predict the future population dynamics of the Atlantic menhaden and striped bass up to the year 2100, which is shown in Figure \ref{predictGraph}. The system converges to equilibrium past 2050, and using the obtained parameters from Table \ref{ParmsTable}, we calculate the equilibrium values as $Q = (x^*, y^*) = (0.1363, 0.2886)$ for the Atlantic menhaden and striped bass respectively. In terms of abundance, this means that the steady state for the Atlantic menhaden is about 136.3 million and the striped bass' is 288.6 million.

\subsection{Resilience of the menhaden-bass system}
Resilience ($R$) measures the stability of a system and is defined as the inverse of the time required for the system to recover from a perturbation and return back to its original steady state. According to Pimm and Lawton \cite{Pimm_Lawton}, the time required for the system to return to equilibrium is given by 
$$
\tau = \displaystyle\frac{-1}{Re(\lambda_m)} \quad \mbox{and} \quad R = \dfrac{1}{\tau},
$$
where $Re(\lambda_m)$ is the real part of the leading eigenvalue of the system, and $\tau$ is time. Note that we can talk about resilience only if the system is stable. As the real part of the leading eigenvalue will be negative, the resilience $\tau$ will be a positive value. As $R$ and $\tau$ are inversely proportional, the higher the resilience of the system, the lower the time for its return to equilibrium when perturbed from its steady state. In our case study with Atlantic menhaden as the prey and striped bass as the predator, we found the eigenvalues of the fitted model to the given observational data corresponding to the co-existence equilibrium to be
\begin{equation}\label{eigenvalues}
\lambda_1 \approx -0.0768 + 0.8301i, \qquad \lambda_2 \approx -0.0768 - 0.8301i.
\end{equation}
As the real part has a negative sign, this means that the co-existence equilibrium is a stable one, and hence, we can calculate the resilience of this system and the return time to equilibrium as
\label{Tau}
$$
R \approx 0.0768 \quad \mbox{and} \quad \tau \approx \frac{1}{0.0768} \approx 13.021.
$$
This means that if the system is perturbed from its steady state it will return to its equilibrium after about 13 years.

\section{Ecological and economic considerations}\label{sec3}
Understanding the interactions between species in the Chesapeake Bay is important, especially when it comes to predator-prey interactions. However, even more important is to understand these interactions when we introduce a human factor in the form of harvesting into the system. Not only are the Atlantic menhaden and the striped bass two very important species in the Chesapeake Bay in terms of their roles in the ecological food chain, but they have proven to be key economic factors as well. Harvesting these species for economic gain makes sense, but when does harvesting becomes excessive and thus detrimental to their populations? How does harvesting affect the predator-prey dynamics? How can we balance harvesting for economic gains and long-term sustainability of these species? What are some policies that could be implemented to achieve such a balance? 

In our study, we try to shed light on some of these important questions. 
We employ mathematical modeling to study the complexity of the problems and the possibility to balance economically effective policies with fisheries conservation for future generations. 

\subsection{Bionomic model}
Let us consider a system where both the prey and the predator are harvested based on the catch per unit effort hypothesis. Incorporating this into the model (\ref{LV_model}) yields to a system of two ordinary differential equations. 
After adding an economic rent function and representing the profit made from harvesting and selling, we get the following bioeconomic, or bionomic, model:
\begin{equation}\label{general_Bionomic_system}
\left\{
\def\arraystretch{2}
\begin{array}{lll}
\displaystyle\frac{dx}{dt} &=& x(r-ax-by)-q_{1}E_{1}x\\
\displaystyle\frac{dy}{dt} &=& y(-e-cy+dx)-q_{2}E_{2}y \\
\pi(E_1, E_2) &=& (p_1q_1x - c_1)E_1 + (p_2q_2y- c_2)E_2 
\end{array}
\right.
\end{equation}
with initial data values $x(0) \geq 0, y(0) \geq 0$. Here, $E_1$ and $E_2$ are the harvesting efforts of the Atlantic menhaden and striped bass respectively, $q_1$ and $q_2$ represent the catchability coefficients of each species. In the right hand side of $\pi$ are the fishing cost per unit effort for both species, $c_1$ and $c_2$, and the price per unit biomass of the species, $p_1$ and $p_2$.

\subsection{Positivity and boundedness of the solutions}\label{sec4}
\begin{figure}[!ht]
    \centering
    \includegraphics[scale=0.5]{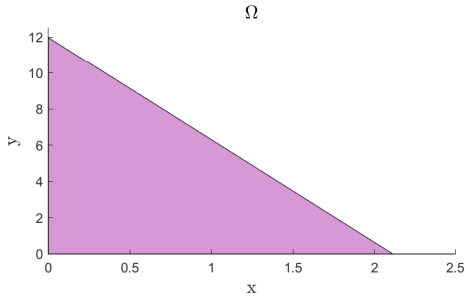}
    \caption{The region $\Omega$ obtained using the parameter values from Table \ref{ParmsTable}. Only the first quadrant is biologically relevant.}
    \label{Omega}
\end{figure}

In this section, we determine a region $\Omega$ where the state variables are positive and bounded. We show that the set is defined as follows
\begin{equation}\label{invariant_set}
\Omega = \left\{ (x,y) \| \quad 0 \leq x \leq \frac{r}{a},  0 \leq x+\frac{y}{\gamma} \leq L \right\},
\end{equation}
for some $\gamma$ and $L$, which will be defined later, is an invariant set, meaning that every solution with initial conditions in this set will remain in it for all $t \geq 0$. Figure \ref{Omega} illustrates the region, using the parameter values from Table \ref{ParmsTable}.

\subsubsection{Positivity of the solutions}
Rewriting equation one of model (\ref{general_Bionomic_system}) in the form
$$
\frac{dx}{x} = \phi_1 dt,
$$
where $\phi_1 (x,y) = r - ax - by$,
and integrating over $[0, t]$ we obtain
$$
x(t) = x(0) \exp{\left[\int_0^y \phi_1 (x,y) ds\right]}.
$$
Since $x(0)\geq 0$, we can conclude that $x(t) \geq 0 $ for all $t\geq 0$.
Similarly, from the second equation of (\ref{general_Bionomic_system}), we have
$$
\frac{dy}{y} = \phi_2 dt,
$$
where $\phi_2 = dx - e - cy$.
Integrating both sides over $[0, t]$, we obtain
$$
y(t) = y(0) \exp{\left[\int_0^y \phi_2 (x,y) ds\right]},
$$
and as the initial condition is positive, $y(0) \geq 0$, then the solution is $y(t) \geq 0$ for all $t \geq 0$, which proves the positivity of the solutions in the defined region $\Omega$.

\subsubsection{Boundedness of the solutions}
Next, we prove that all solutions remain bounded in $\Omega$, which means that as long as the initial conditions are in $\Omega$, the solutions to (\ref{general_Bionomic_system}) remain in $\Omega$ for all $t$. 

As all coefficients are non-negative, we have $\dfrac{dx}{dt} = x(r-ax-by)-q_{1}E_{1}x \leq x(r-ax)$. 
Now solving $\dfrac{dx}{dt} = x(r-ax)$ by separation of variables and using the partial fractions decomposition method, we get:
$$
    \left(\frac{1}{rx} + \frac{a}{r} \frac{1}{r-ax}\right){dx} = {dt}. 
$$
After integrating both sides and some manipulations, we can express $x$ explicitly as:
$$
x = \frac{Cr e^{rt}}{1+aC e^{rt}} = \frac{Cr}{aC +e^{-rt}},
$$
where $C$ is a constant of integration. 
Now letting $t \rightarrow \infty, \text{implies that } x \rightarrow  \frac{r}{a}.$ Therefore $x(t) \leq \frac{r}{a}$ is bounded in $\Omega$.

Now we want to show that $y(t)$ is also bounded as $t \rightarrow \infty$. To prove that, we introduce a new variable, $w(t)$, defined as
\begin{equation}\label{W(t)_ybound}
    w(t) = x(t) + \frac{y(t)}{\gamma},
\end{equation}
where $\gamma$ is given by $\gamma = \frac{d}{b}$ or equivalently $d = \gamma b$. Using the first two equations in (\ref{general_Bionomic_system}), we obtain the following differential equation for the new variable $w$
$$ \frac{dw}{dt} = \frac{dx}{dt} + \frac{1}{\gamma}\frac{dy}{dt} = x(r-ax) - q_{1}E_{1}x - \frac{1}{\gamma}(ey+cy^{2}+q_{2}E_{2}y).
$$
Using the fact that $-cy^{2}<0$, we find the differential inequality:
$$
    \frac{dw}{dt} \leq x(r-ax) - q_{1}E_{1}x - \frac{y}{\gamma}(e+q_{2}E_{2}).
$$
Observe that if we let $f(x) = x(r-ax)$, then $f(x)$ obtains its maximum at $x=\dfrac{r}{2a}$ and that maximum is $f(\dfrac{r}{2a}) = \dfrac{r^{2}}{4a}$. Substituting this back into the differential inequality, along with letting $\delta = \min{(q_{1}E_{1},q_{2}E_{2},e)}$, we get:
\begin{align*}
    \frac{dw}{dt} &\leq x(r-ax) - q_{1}E_{1}x - \frac{y}{\gamma}(e+q_{2}E_{2})\\
    &\leq \frac{r^{2}}{4a} - \delta (x+\frac{y}{\gamma})
    = \frac{r^{2}}{4a} - \delta w.
\end{align*}
Thus, the new differential inequality for $w$ is as follows
$$
\frac{dw}{dt} + \delta w \leq \frac{r^{2}}{4a}.
$$
Using the integrating factor, multiplying both sides by $e^{\delta t}$, and integrating, after some manipulations we get 
$$
w(t) \leq \frac{w(0)}{e^{\delta t}} - \frac{r^{2}}{4a\delta e^{\delta t}} + \frac{r^{2}}{4a\delta} = \frac{1}{e^{\delta t}}(w(0) - \frac{r^{2}}{4a\delta}) + \frac{r^{2}}{4a\delta}.
$$

From here, as $t\rightarrow \infty$, $e^{\delta t}$ $\rightarrow \infty$, which in turn means that the term $\frac{1}{e^{\delta t}}(w(0) - \frac{r^{2}}{4a\delta}) \rightarrow{0}$ as $t \rightarrow \infty$, reducing the above inequality to
$$
\lim_{t\to\infty}w(t) \leq \frac{r^{2}}{4a\delta}.
$$
From here, $w = x + \frac{y}{\gamma}$ is bounded and hence are solutions in the region $\Omega$, where $L = \frac{r^{2}}{4a\delta}$. Therefore, the region $\Omega$ is an invariant set, and any solution that has initial conditions in $\Omega$ remains in $\Omega$. With this, all solutions of the model (\ref{general_Bionomic_system}) are nonnegative and bounded in $\Omega$.

\subsection{Bionomic equilibrium existence}
The bionomic equilibrium is obtained when $dx/dt = dy/dt =0$ and when the total revenue earned by selling the harvested biomass is equal to the total cost for the effort in harvesting the biomass.
Here, we want to find conditions under which bionomic equilibria exist. Suppose the bionomic equilibria is $P(x_{B}, y_{B}, E_{1B}, E_{2B})$. Then $P$ will satisfy the system
\begin{equation}\label{Bionomic_equilibrium_system}
\left\{
\def\arraystretch{1.2}
\begin{array}{rrr}
r-ax-by-q_{1}E_{1} &=& 0\\
-e-cy+dx-q_{2}E_{2} &=& 0 \\
(p_1q_1x - c_1)E_1 + (p_2q_2y- c_2)E_2 &=& 0
\end{array}
\right.
\end{equation}
Note that we did not consider the biological equilibrium $E_{0}(x, y) = (0, 0)$ because the overall net revenue $\pi$ becomes negative. To find a solution to (\ref{Bionomic_equilibrium_system}), we will consider four different cases based on the harvesting efforts.

\subsubsection*{Case 1. Harvesting the predator only ($E_1 =0$, $E_2 \ne 0$)}
If the fishing cost per unit effort for the prey (menhaden) is greater than the revenue obtained from this fishing effort, $c_1 > p_1 q_1 x$, the fishermen will naturally withdraw from this fishery and as a result the fishing effort for menhaden will be zero $E_1 = 0$, but the effort for striped bass will be not zero $E_2 \neq 0 $. Then for the third equation in (\ref{Bionomic_equilibrium_system}) to be true, we can take $y_B = c_2/p_2 q_2$.
From the first equation of (\ref{Bionomic_equilibrium_system}) we can express $x_B$:
$$
x_B = \displaystyle\frac{r}{a} - \frac{b c_2}{a p_2 q_2},
$$
while from the second equation we can express $E_{2B}$ as:
\begin{equation}\label{Case-1-E_2B}
E_{2B} = - \displaystyle\frac{e}{q_2} - \frac{c c_2}{p_2 q_2^2} + \frac{d}{q_2}\left(\frac{r}{a} - \frac{b c_2}{a p_2 q_2}\right).
\end{equation}

\begin{theorem}
If there are no harvesting efforts on the prey, the bionomic equilibrium $$P_1\left(x_B, y_B, E_{1B}, E_{2B}\right) = P_1\left(\displaystyle\frac{r}{a} - \frac{b c_2}{a p_2 q_2}, \frac{c_2}{p_2 q_2}, 0, E_{2B}\right),$$ where $E_{2B}$ is given by (\ref{Case-1-E_2B}), exists provided that the following two conditions hold:
$$
 \displaystyle\frac{r}{a} > \frac{b c_2}{a p_2 q_2}
$$
$$
 \frac{d}{q_2}\left(\frac{r}{a} - \frac{b c_2}{a p_2 q_2} \right) >  \displaystyle\frac{e}{q_2} + \frac{c c_2}{p_2 q_2^2}.
 $$
\end{theorem}

\subsubsection*{Case 2. Harvesting the prey only ($E_1 \ne 0$, $E_2 =0$)} 
Likewise, if the fishing cost per unit effort for the predator (striped bass) is greater than the revenue obtained from this fishing effort, $c_2 > p_2 q_2 x$, the fishermen will naturally withdraw from this fishery and as a result the fishing effort for the striped bass will be zero $E_2 = 0$, but the effort for Atlantic menhaden will be not zero $E_1 \neq 0 $.
Then for the third equation in (\ref{Bionomic_equilibrium_system}) to be true, we can take $x_B = c_1/p_1 q_1$.
From the second equation of (\ref{Bionomic_equilibrium_system}) we can express $y_B$:
$$
y_B = \displaystyle\frac{d c_1}{c p_1 q_1} - \frac{e}{c},
$$
while from the first equation we can express $E_{1B}$  as
\begin{equation}\label{Case-1-E_1B}
E_{1B} = \displaystyle\frac{r}{q_1} - \frac{a c_1}{p_1 q_1^2} - \frac{b}{q_1}\left(\frac{d c_1}{c p_1 q_1} - \frac{e}{c}\right).
\end{equation}

\begin{theorem}
If there are no harvesting efforts on the predator, the bionomic equilibrium $$P_2\left(x_B, y_B, E_{1B}, E_{2B}\right) = P_2\left(\frac{c_1}{p_1 q_1}, \displaystyle\frac{d c_1}{c p_1 q_1} - \frac{e}{c}, E_{1B}, 0\right),$$ where $E_{1B}$ is given by (\ref{Case-1-E_1B}), exists provided that the following two conditions hold:
$$
 \displaystyle\frac{d c_1}{c p_1 q_1} > \frac{e}{c}
$$
$$
  \displaystyle\frac{r}{q_1} - \frac{a c_1}{p_1 q_1^2} >  \frac{b}{q_1}\left(\frac{d c_1}{c p_1 q_1} - \frac{e}{c}\right).
 $$
\end{theorem}

\subsubsection*{Case 3. No harvesting of both the prey and the predator ($E_1 = E_2 = 0$)}
If $c_1 > p_1 q_1x$ and $c_2 > p_2 q_2y$, then the fishing cost is greater than the revenue for menhaden and striped bass and the fisheries will be closed. We can not have a bionomic equilibrium in this case, however the biological equilibrium $(x^*, y^*)$ will still exist if $rd > ae$:
$$
x^* = \frac{rc + be}{ac + bd}, \qquad y^* = \frac{rd - ae}{ac + bd}.
$$

\subsubsection*{Case 4. Simultaneous harvesting of both the prey and the predator ($E_1 \neq0, E_2 \neq 0$)}

If $c_1 \leq p_1q_1x$ and $c_2 \leq p_2q_2y$, then the cost of fishing is less or equal to the revenue for both the menhaden and the striped bass, and the fisheries have an incentive to harvest both species, and thus $E_1 \neq 0$ and $E_2 \neq 0$. Therefore, to find the bionomic equilibrium of (\ref{Bionomic_equilibrium_system}) and in order for the third equation to be true, we can set $x_B$ and $y_B$ as
$$
x_B = \frac{c_1}{p_1q_1}, \quad y_B = \frac{c_2}{p_2q_2}
$$
respectively. From these, we can express $E_1$ and $E_2$ from the first and second equations in (\ref{Bionomic_equilibrium_system}) respectively:

\begin{equation}\label{Case-4_E1_E2}
\begin{array}{ccc}
E_{1B} &=& \displaystyle\frac{r}{q_1} - \frac{ac_1}{p_1q_1^2} - \frac{bc_2}{p_2q_1q_2}\\
E_{2B} &=& \displaystyle\frac{-e}{q_2} - \frac{cc_2}{p_2q_2^2} + \frac{dc_1}{p_1q_1q_2}.
\end{array}
\end{equation}

\begin{theorem}
If both the prey and the predator are harvested, the bionomic equilibrium $$P_3\left(x_B, y_B, E_{3B}, E_{3B}\right) = P_1\left(\displaystyle\frac{c_1}{p_1q_1}, \frac{c_2}{p_2 q_2}, E_{1B}, E_{2B}\right),$$ where $E_{1B}$ and $E_{2B}$ are given by (\ref{Case-4_E1_E2}), exists provided that the following two conditions hold:
$$
 \displaystyle\frac{r}{q_1} > \frac{ac_1}{p_1q_1^2} + \frac{bc_2}{p_2q_1q_2}
$$
$$
 \displaystyle\frac{dc_1}{p_1q_1q_2} > \frac{e}{q_2} + \frac{cc_2}{p_2q_2^2}.
 $$
\end{theorem}

\subsection{Effect of harvesting on the stability of the predator-prey system}
To study the stability of the biological system when either the prey, or the predator, or both prey and predator are harvested we will use the Jacobian of the system (\ref{Bionomic_equilibrium_system}):
\begin{equation}\label{Jacobian}
J(x,y) = 
\left(
\begin{array}{ccc}
r - 2ax -by - q_1E_1 & -bx \\
dy & -e - 2cy + dx - q_2 E_2 
\end{array}
\right).
\end{equation}

\subsubsection*{\textbf{Case 1. Harvesting the predator only}}
If only the predator is harvested, the co-existence equilibrium $(\hat{x}, \hat{y})$ is the solution to the system:
\begin{equation}
\left\{
\begin{array}{ccc}
ax+by &=& r \\
dx-cy &=& e+q_2 E_2. \\
\end{array}
\right.
\end{equation}
Using Cramer's rule, the solutions are
\begin{equation}\label{Case1_x_hat}
    \hat{x} = \displaystyle\frac{b(e+q_2E_2) + rc}{ac + bd},
\end{equation}

\begin{equation}\label{Case1_y_hat}
    \hat{y} = \displaystyle\frac{rd - a(e+q_2E_2)}{ac + bd}.
\end{equation}
Note that $\hat{x}$ is always positive; hence, for the co-existence equilibrium to exist, we require $\hat{y} > 0$, which provides a constraint on the harvesting effort for the predator:
\begin{equation}\label{Case1_E_2_inequality}
 0 \le E_2 < \frac{rd - ae}{aq_2}.
\end{equation}

Substituting $(\hat{x}, \hat{y})$ into the Jacobian matrix (\ref{Jacobian}) and using the properties of the equilibrium solution, we get the simplified Jacobian matrix:
\begin{equation}\label{Jacobian1}
J(\hat{x}, \hat{y}) = 
\left( 
\begin{array}{ccc}
-a\hat{x}  & -b \hat{x} \\
d \hat{y} & - c \hat{y}  
\end{array}
\right) = A.
\end{equation}
Its characteristic equation is the quadratic equation
$$
\lambda^2 - trace(A) \lambda + det(A) = 0,
$$
where $trace(A) = -(a \hat{x} + c \hat{y})$ and $det(A) = (ac + bd)\hat{x} \hat{y}$. Note that the trace is always negative, while the determinant is always positive, which implies that the eigenvalues will have negative real parts. Hence, the system is asymptotically stable for any harvesting effort $E_2$ that satisfies condition (\ref{Case1_E_2_inequality}). 

To obtain the upper bound of $E_2$ from (\ref{Case1_E_2_inequality}), we need to choose a value for the catchability parameter $q_2$. The range for catchability parameters is $[0, 1]$, with lower values used when the fish is relatively easy to catch, such as forage fish, and higher values when the fish is more rare or difficult to catch. Here, as $q_2$ is the catchability parameter for striped bass, without loss of generality we choose $q_2 =1$. Then using the parameters from Table \ref{ParmsTable}, we find that the co-existence equilibrium for Atlantic menhaden and striped bass is asymptotically stable as long as the effort is $E_2 \lessapprox 198.4365$. 

\begin{figure}[!ht]
    \centering
    \includegraphics[width=\textwidth]{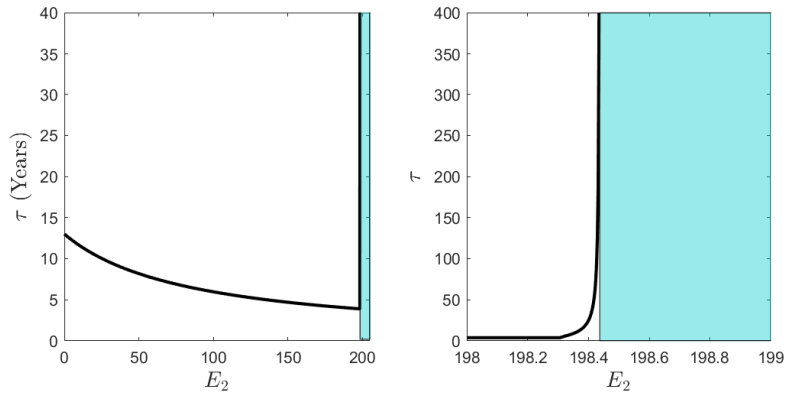}
    \caption{Return time to equilibrium $\tau$ versus the effort $E_2$ when harvesting the predator only. The panel to the left shows the return time to equilibrium for the full range $0 \le E_2 \le 198.4365$; the panel to the right shows the return time to equilibrium for $198 \le E_2 <199$.}
    \label{E2_tau}
\end{figure}

To analyze the resilience of the menhaden-bass system, we calculate the return time to equilibrium shown in Figure \ref{E2_tau}. The return time of the menhaden-bass system to equilibrium decreases from 13 years as a function of $E_2$ up until $E_2 \approx 198.3$, and approaches about 3 years close to the upper end of the interval. For $E_2 \gtrsim 198.3$, the return time to equilibrium increases significantly and eventually grows asymptotically towards infinity as $E_2$ approaches the upper bound of $\sim198.4365$.

\subsubsection*{\textbf{Case 2. Harvesting the prey only}}
If only the predator is harvested, the co-existence equilibrium $(\bar{x}, \bar{y})$ is the solution to the system:
\begin{equation}
\left\{
\begin{array}{ccc}
r-ax-by &=& q_1 E_1 \\
-e-cy+dx &=& 0 \\
\end{array}
\right.
\end{equation}
Using Cramer's rule, the solutions are
\begin{equation}\label{Case2_x_bar}
\bar{x} = \displaystyle\frac{c(r-q_1E_1) + be}{ac + bd},
\end{equation}
\begin{equation}\label{Case2_y_bar}
\bar{y} = \displaystyle\frac{d(r-q_1E_1) - ae}{ac + bd}.
\end{equation}
Note that $\bar{x} > 0$ requires that $r > q_1E_1$, or equivalently, $E_1 < \displaystyle\frac{r}{q_1}$ which is a natural condition for existence of the Atlantic menhaden, as otherwise, if $r < q_1E_1$, this means that the Atlantic menhaden are being harvested to extinction without predation from the bass. Now for the co-existence equilibrium to be biologically relevant, we require $\bar{y} > 0$, which provides a constraint on the prey's harvesting effort:
\begin{equation}\label{Case2_E_1_inequality}
   0 \le E_1 < \frac{r}{q_1} - \frac{ae}{dq_1} = \dfrac{rd - ae}{dq_1}.
\end{equation}
Observe also that from the condition (\ref{Case2_E_1_inequality}), it follows that $E_1 < \displaystyle\frac{r}{q_1}$ and hence $\bar{x} >0$.  
Substituting $(\bar{x}, \bar{y})$ into the Jacobian matrix (\ref{Jacobian}) and using the properties of the equilibrium solution, we get the simplified Jacobian matrix:
\begin{equation}\label{Jacobian2}
J(\bar{x}, \bar{y}) = 
\left( 
\begin{array}{ccc}
-a\bar{x}  & -b \bar{x} \\
d \bar{y} & - c \bar{y}  
\end{array}
\right) = \bar{A}.
\end{equation}
Similar considerations about the trace and determinant of $\bar{A}$ as before lead to the conclusion that  the coexistence equilibrium $(\bar{x}, \bar{y})$ is asymptotically stable provided condition (\ref{Case2_E_1_inequality}) for the prey effort $E_1$ is satisfied. 

To obtain the upper bound of the harvesting effort $E_1$ from (\ref{Case2_E_1_inequality}), we need to choose a value for the catchability parameter $q_1$. Since the Atlantic menhaden is a forage fish that travels in schools, we assume that it is relatively easy to catch and hence $q_1 = 0.5 << 1$. Then, using the model's parameters from Table \ref{ParmsTable}, this gives that the equilibrium solutions $(\bar{x}, \bar{y})$ are asymptotically stable as long as the effort on menhaden is bounded by $E_1 \lessapprox 1.0198$. 

\begin{figure}[!ht]
    \centering
    \includegraphics[width=\textwidth]{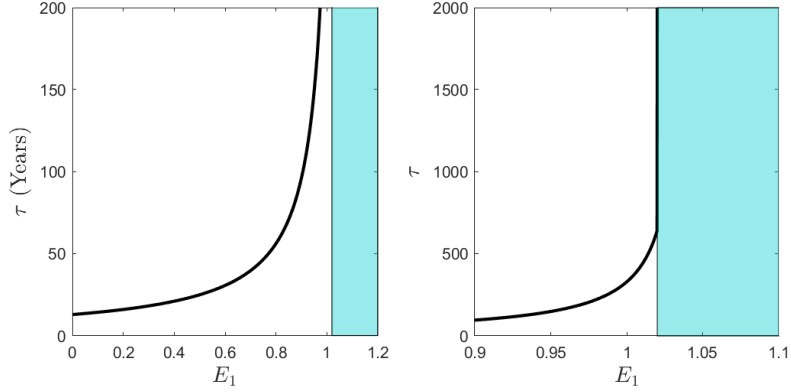}
    \caption{Return time to equilibrium $\tau$ versus the effort $0 \le E_1 < 1.2$ when harvesting the prey only. The panel to the right is zoomed in when the harvesting effort is $0.9 \le E_1 < 1.1$.}
    \label{E1_tau}
\end{figure}

To analyze the resilience of the menhaden-bass system, we calculate the return time to equilibrium shown in Figure \ref{E1_tau}.  The return time to equilibrium increases monotonically as a function of $E_1$ and eventually grows asymptotically towards infinity as $E_1$ approaches the upper bound of $\sim 1.0198$.

\subsubsection*{\textbf{Case 3. Combined harvesting of both the prey and the predator}} \label{3.4 case 3}
Let us now apply a combined harvesting effort $E_1 = E_2 = E$ to both the prey and the predator. The coexistence equilibrium $(x^*, y^*)$ is a solution to the system:
\begin{equation}
\left\{
\begin{array}{ccc}
ax+by &=& r-q_1E \\
dx-cy &=& e+q_2E. \\
\end{array}
\right.
\end{equation}
Here, the solutions are
\begin{equation}\label{Case3_x_hat}
x^* = \displaystyle\frac{(cr + be) - (cq_1 - bq_2)E}{ac + bd},
\end{equation}
\begin{equation}\label{Case3_y_hat}
y^* = \displaystyle\frac{(rd- ae) - (dq_1 + a q_2) E}{ac + bd}.
\end{equation}
Therefore, in order for the coexistence equilibrium to exist, we require $x^* >0$ and $y^* >0$. We first consider $x^*$. If $cq_1 > bq_2$, we have the following constraint on the catchability coefficients:
$$
\dfrac{q_1}{q_2} > \dfrac{b}{c} \approx 3.4,  \quad \mbox{or} \quad q_1 > 3.4 q_2.
$$
This condition means that it is three times harder to catch menhaden than the bass, which is not a realistic assumption due to the Atlantic menhaden being a forage fish which travel in schools. Thus, we only consider the case when $cq_1 < bq_2$, which forces $x^*>0$ for positive effort.
Requiring $y^* > 0$ provides the following constraint on the effort:
\begin{equation}\label{Combined_Harvesting_Effort}
0 \le E < \displaystyle\frac{rd - ae}{dq_1 + aq_2}.
\end{equation}
Substituting $(\hat{x},\hat{y})$ into the Jacobian matrix, we get the following:
$$
J(x^*, y^*) = 
\left(
\begin{array}{ccc}
r - 2a x^* -b y^* - q_1E  & -b x^* \\
dy^* & -e - 2cy^* + dx^* - q_2 E 
\end{array}
\right) = 
\left( 
\begin{array}{ccc}
-ax^*  & -b x^* \\
d y^* & - c y^*  
\end{array}
\right) = A^*.
$$ 
Therefore, the coexistence equilibrium $(x^*, y^*)$ is asymptotically stable as long as the condition (\ref{Combined_Harvesting_Effort}) for the combined harvesting effort is satisfied.

Using our obtained parameters and assuming $q_1 = 0.5, q_2 = 1$, we get that the equilibrium is asymptotically stable as long as $E < \displaystyle\frac{rd - ae}{dq_1 + aq_2} \approx 1.0146$.  The return time to equilibrium increases monotonically as a function of $E$ and eventually grows asymptotically towards infinity as $E$ approaches the upper bound of $\sim 1.0146$.

\begin{figure}[!ht]
    \centering
    \includegraphics[width=\textwidth]{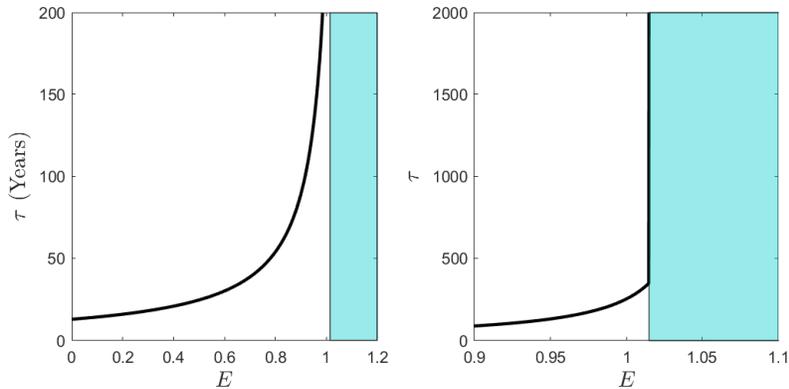}
    \caption{Return time to equilibrium $\tau$ versus the effort $0 \le E < 1.2$ when harvesting both the prey and the predator with a same effort $E$. The panel to the right is zoomed in when the harvesting effort is $0.9 \le E < 1.1$.}
    \label{E_tau}
\end{figure}

\section{Analysis of harvesting policies}

\subsection{Maximum sustainable yield}
We now want to analyze how existing fishery management policies might affect the biological sustainability and economic gain within our two-species bionomic system (\ref{general_Bionomic_system}). We will first consider the implementation of a maximum sustainable yield (MSY) policy on our system. MSY is a theoretical management approach that is defined as the highest catch of a species that maximizes the yield  while simultaneously ensuring that the species population is at a sustainable level \cite{fathEcoBook}. The concept of MSY originated in Schaefer's 1954 paper \cite{schaeferPaper} and has since been widely considered and researched \cite{msyPaper8, msyPaper3, msyPaper5, msyPaper2}. Even though MSY is one of the most frequently used policies for fisheries management, it has also been strongly - and rightly - challenged by many scientists \cite{Finley, Holt-2011, Holt-2018}.

Here, we will apply MSY to our bionomic system as it allows for a conceptually easy comparison to other policies that we consider. In addition, we will study the combined application of MSY principles and ecological based models on the long term fish population sustainability. As we wish to harvest both fishes such that we can get maximum yield sustainably, our goal is to maximize the yield function $Y(E) = q_1E x^* + q_2 E y^*$ with respect to $E$ when both species are harvested with the same effort $E$. Here, $(x^*, y^*)$ is the equilibrium point of the predator-prey system.

\begin{figure}[!ht]
    \centering
    \includegraphics[scale=0.6]{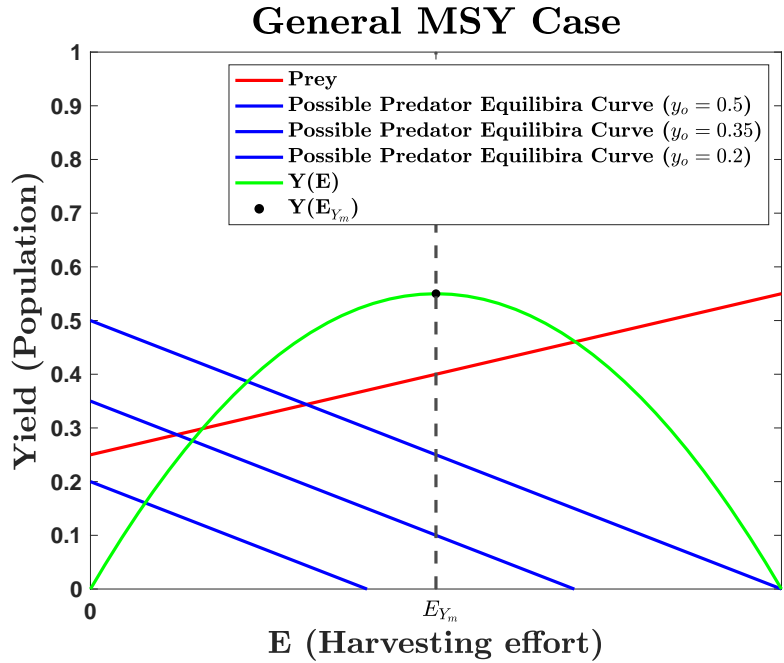}
    \caption{A general case for two species MSY analysis with multiple predator equilibria lines. }
    \label{Figure_General_Case}
\end{figure}

To illustrate how we will apply MSY policy in this study, let us use Figure \ref{Figure_General_Case}. The yield function is given in green and the prey and predator equilibria curves, which are linearly dependent on effort, are given in red and blue respectively. We will denote $E_{MSY}$ the effort level needed to achieve maximum sustainable yield (MSY). Note that to be ``sustainable", this yield should guarantee that both species are sustainable, meaning that we require both $x>0$ and $y>0$. 
As the yield function is a parabola, it reaches its maximum when the effort is at $E_{Y_m}$, which is marked with a dashed line in Figure \ref{Figure_General_Case}.
If the predator equilibrium curve is similar to the top blue line, then the effort level corresponding to the maximum of the yield curve, $E_{Y_m}$, applied to both species will result in a sustainable predator-prey system; hence, such an effort is the desirable $E_{MSY}$. However, if the predator curve follows the middle blue line, then harvesting at $E_{Y_m}$ level both the prey and the predator will result in a positive, but near 0, equilibrium population of the predator. This means that while still ``sustainable", the predator population is approaching quite low level, which may result in higher existential risk due to the practical difficulties of low populations in reality. And lastly, if the predator's equilibrium curve follows the bottom blue line, harvesting at effort close to $E_{Y_m}$ value will result in extinction of the predator's population. Therefore, to be sustainable, the maximum possible effort allowed should be much lower than the effort maximizing the yield; hence, MSY could be archived but with an effort that is less than the effort maximizing the yield. 

Next, we will consider each of the three scenarios: harvesting the predator only, harvesting the prey only, and combined harvesting of both the prey and the predator within our particular two-species model. 

\subsubsection*{Case 1. Harvesting the predator only}
Let us assume that only the predator is harvested with an effort $E_2 = E$. As we already calculated the equilibrium values $(\hat{x}, \hat{y})$ in (\ref{Case1_x_hat}) and (\ref{Case1_y_hat}), the yield at the equilibrium as a function of the effort is defined as follows
\begin{equation}\label{yield_predator_harvesting}
Y(E) = E \hat{y}(E) = E \left(\displaystyle\frac{rd -ae}{ac + bd} - \frac{aq_2}{ac + bd}E \right).
\end{equation}
As the yield function is a quadratic function with respect to $E$, we can easily find the effort at which the yield attains its maximum, namely $E_{Y_m} = \displaystyle\frac{rd - ae}{2aq_2}$.
Next, we can calculate both $\hat{x}$ and $\hat{y}$ with respect to $E$ as follows
$$
\hat{x}(E) = \displaystyle\frac{be + rc}{ac + bd} + \frac{bq_2}{ac + bd}E; \quad
\hat{y}(E) = \displaystyle\frac{rd -ae}{ac + bd} - \frac{aq_2}{ac + bd}E,
$$
where both $\hat{x}$ and $\hat{y}$ are linearly dependent on $E$. Observe that as $E$ increases, $\hat{x}$ is increasing as the slope of the line is positive, and $\hat{y}$ is decreasing as the slope of the line is negative.

\begin{figure}[!ht]
\centering
\includegraphics[width=5.15in, scale=0.6]{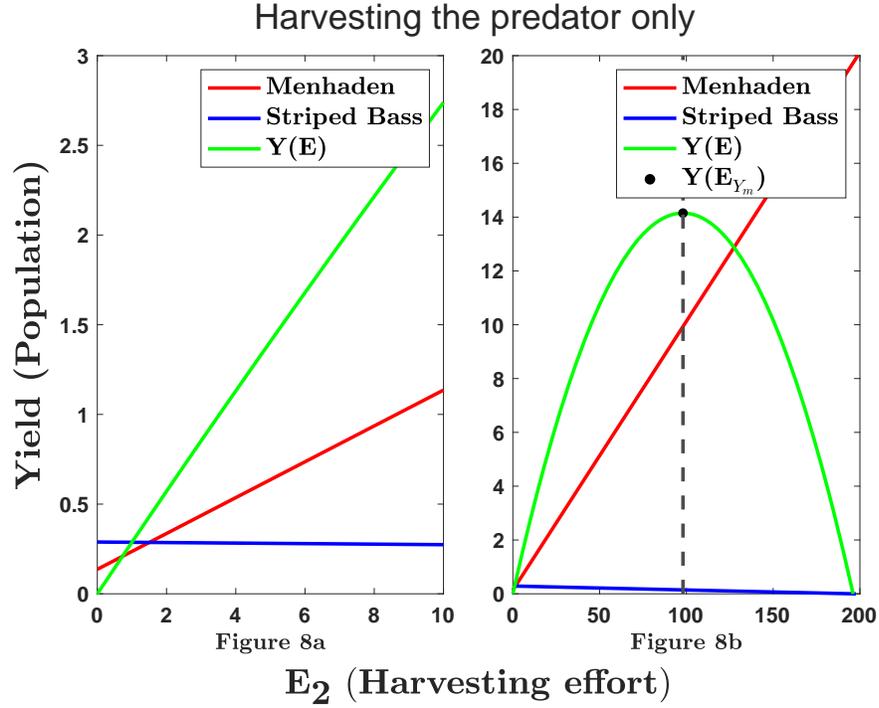}
\caption{Yield function (in green) and equilibria curves for menhaden (in red) and bass (in blue) when only the predator is harvested using the parameters from Table \ref{ParmsTable}, with $q_2 = 1$. The panel to the left is zoomed in when the harvesting effort is given by the range $0 \le E_2 \le 10$.}
\label{Figure_Predator_Harvesting}
\end{figure}

Note that harvesting the predator only reduces the population of the predator but the prey's population increases  at equilibrium. 
Assuming that the predator's harvesting is done at level $E_{Y_m}$, we can calculate  $\hat{x}(E_{Y_m}) = \displaystyle\frac{be + rc}{ac + bd} + \frac{bq_2}{ac + bd}E_{Y_m}$ and
$\ \hat{y}(E_{Y_m}) = \displaystyle\frac{rd-ae}{2(ac + bd)}$.

Hence, the maximum yield will be sustainable if both $\hat{x}(E_{Y_m})$ and $\hat{y}(E_{Y_m})$ are positive, which will be true if $rd > ae$ is satisfied. Therefore, maximum sustainable yield could be achieved by harvesting at $E_{Y_m}$ level, with corresponding equilibrium values of $\hat{x}(E_{MSY}) = \hat{x}(E_{Y_m})$ and $\hat{y}(E_{MSY}) = \hat{y}(E_{Y_m})$ when $rd > ae$ for model (\ref{general_Bionomic_system}). The predator's population in this scenario is half of its population in the absence of harvesting. We may state the following theorem about MSY in this case.

\begin{theorem}
\label{theorem_pred_harv}
When harvesting the predator only in the system (\ref{general_Bionomic_system}), the MSY will exist and will be achieved with an effort $E_{MSY} = E_{Y_m} = \displaystyle\frac{rd - ae}{2aq_2}$ if $rd > ae$.
\end{theorem}

For our particular model with parameters from Table \ref{ParmsTable}, the case of harvesting the predator only is illustrated in Figure (\ref{Figure_Predator_Harvesting}). It is easy to check that Theorem \ref{theorem_pred_harv} is satisfied as  $rd \approx 5.1303$ and $ae \approx 0.0312$, and thus $rd > ae$. For the menhaden-bass system under consideration, harvesting the striped bass at the MSY level can occur. 

\subsubsection*{Case 2. Harvesting the prey only}
Next, let us consider the case when only the prey is harvested with an effort $E_1 = E$. For the equilibrium $(\hat{x}, \hat{y})$ from (\ref{Case2_x_bar}) and (\ref{Case2_y_bar}), the yield at equilibrium as a function of effort is defined as follows:
\begin{equation}\label{yield_prey_harvesting}
Y(E) = E \hat{x}(E) = E \left(\displaystyle\frac{cr + be}{ac + bd} - \frac{cq_1}{ac + bd}E \right).
\end{equation}
As the yield function is a quadratic function with respect to $E$, the effort where the parabola reaches its maximum is given as $E_{Y_m} =  \displaystyle\frac{cr + be}{2cq_1}$. Next, we calculate both $\hat{x}$ and $\hat{y}$ with respect to $E$:
$$
\hat{x}(E) = \displaystyle\frac{cr + be}{ac + bd} - \frac{cq_1}{ac + bd}E; \quad
\hat{y}(E) = \displaystyle\frac{rd -ae}{ac + bd} - \frac{dq_1}{ac + bd}E,
$$
where both $\hat{x}$ and $\hat{y}$ are linearly dependent on $E$, and both $\hat{x}$ and $\hat{y}$ are decreasing as $E$ increases. Hence, for the MSY to exist, both species must be sustainable at $E_{MSY}$ or $\hat{x}(E_{MSY}) > 0$ and $\hat{y}(E_{MSY}) > 0$ .
Setting $\hat{x}(E)>0$ gives the following restriction on the effort $E < \dfrac{cr + be}{c q_1}$, and setting  $\hat{y}(E) > 0$ gives $E < \dfrac{rd - ae}{d q_1}$.
If we assume harvesting at level $E_{Y_m}$ where the yield reaches its maximum, for both $x$ and $y$ to stay positive, we require that $\dfrac{cr + be}{2c q_1} < \dfrac{cr + be}{c q_1}$ and $\dfrac{cr + be}{2c q_1} < \dfrac{rd - ae}{d q_1}$. The former inequality is always true, and for the second to be true, the following requirement on the system's parameters must be satisfied
$$
\dfrac{rd - ae}{cr+be} > \dfrac{d}{2c}.
$$
Therefore, we may state the following theorem about MSY.
\begin{theorem}
\label{theorem_prey_harv}
If $\dfrac{rd - ae}{cr+be} > \dfrac{d}{2c}$ in the system (\ref{general_Bionomic_system}), MSY exist and could be achieved as a result of harvesting the prey with effort level $E_{MSY} = E_{Y_m} = \displaystyle\frac{cr + be}{2cq_1}$. 
\end{theorem}

\begin{figure}[!ht]
    \centering
    \includegraphics[scale=0.6]{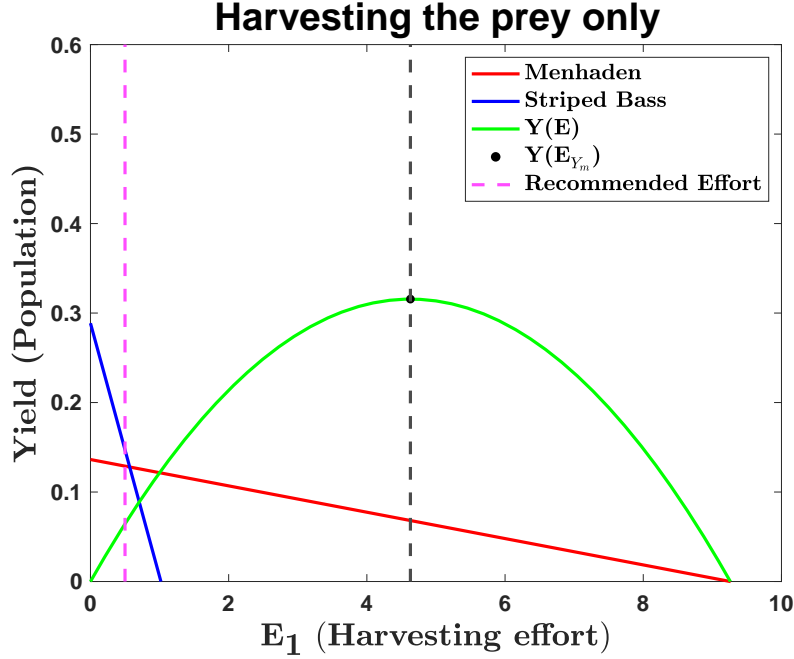}
    \caption{Yield function (in green) and equilibria curves for menhaden (in red) and bass (in blue) when only the prey is harvested, using the parameters from Table \ref{ParmsTable}, with $q_1 = 0.5$. The vertical dashed line in black represents the effort that maximizes the ; the black dot is the maximum yield $Y(E_{Y_m})$. The vertical dashed line in magenta represents the recommended MSY effort. }
    \label{Figure_Prey_Harvesting}
\end{figure}

For the menhaden-bass system, the condition of Theorem \ref{theorem_prey_harv} is not satisfied, and we will hence look for a different effort that returns positive yield while keeping the fish populations sustainable. Figure (\ref{Figure_Prey_Harvesting}) shows the yield function and the equilibria curves for menhaden and bass using the parameters from Table \ref{ParmsTable}. We can see that increasing prey harvesting reduces the populations of both prey and predator at equilibrium. Even more, harvesting at level equal to $E_{Y_m}$ is not sustainable as it results in the extinction of the predator; thus, the sustainable harvesting level must be the level at which the predator does not go extinct: $E_{MSY} <  \displaystyle\frac{rd - ae}{dq_1}$. Hence, for the menhaden-bass system under consideration when harvesting Atlantic menhaden only, to achieve sustainability of both species, the harvesting must be bound as $E_{MSY} < \dfrac{rd - ae}{d q_1} \approx 1.02$. 

Even though harvesting close to this upper limit $E_{max}=\dfrac{rd - ae}{d q_1} \approx 1.02$ keeps the predator at a positive population value, the population numbers are extremely low and the predator species is at a high extinction risk. It is necessary to reduce the allowed effort in order to avoid possible extinction of the predator. One way to do that is by reducing the effort to 50\% $E_{max}$, meaning that the suggested MSY level then becomes $E_{MSY} = \dfrac{rd - ae}{2d q_1} = 0.5$, which is illustrated by the magenta line in Figure(\ref{Figure_Prey_Harvesting}). Similar techniques were reported in  \cite{review-2021-MSY-menhaden} and \cite{balancing_model_complexity_ERP}, where the fishing recommendation was set at 50\%-75\% of the calculated $E_{MSY}$ for Atlantic menhaden. 

These results emphasize the usefulness of multi-species models and the need for addressing the role of forage fish in the ecosystem as food for predators when managing the fisheries. Recently, ecological reference points were introduced by ASFMC to address such need. In \cite{combining_ecosystem_and_singe_species_modeling} it was reported that to address those needs, the ecological reference point target had to be set 40\% lower than the target obtained using a single species model.

\subsubsection*{Case 3. Combined harvesting of both the prey and the predator}

Let us assume combined harvesting of both the prey and the predator with effort $E$. For our equilibrium $(\hat{x},\hat{y})$ from (\ref{Case3_x_hat}) and (\ref{Case3_y_hat}), defining the yield at equilibrium as a function of effort follows
\begin{equation}\label{yield_combined_harvest}
Y(E) = E (q_1\hat{x} + q_2\hat{y}) = E \left(\displaystyle\frac{cr + be + rd - ae}{ac + bd} - \frac{cq_1 - bq_2 + dq_1 + aq_2}{ac + bd}E \right)
\end{equation}
where $E$ is the effort level of harvesting for both the prey and predator. As $Y(E)$ is quadratic, $E_{Y_m} = \displaystyle\frac{cr + be + rd - ae}{2(cq_1 - bq_2 + dq_1 + aq_2)}$. Next, we can calculate both $\hat{x}$ and $\hat{y}$ with respect to E as follows
$$
\hat{x}(E) = \displaystyle\frac{cr + be}{ac + bd} - \frac{cq_1 - bq_2}{ac + bd}E,
$$
$$
\hat{y}(E) = \displaystyle\frac{rd -ae}{ac + bd} - \frac{dq_1 + aq_2}{ac + bd}E,
$$
where both $\hat{x}$ and $\hat{y}$ are linearly dependent on $E$, with $\hat{y}$ is decreasing as the effort increases. For $\hat{x}$, there are two cases, depending on the values of $cq_1$ and $bq_2$.  In the first case, where $cq_1 > bq_2$, $\hat{x}(E)$ is decreasing. In the second case, when $cq_1 < bq_2$, $\hat{x}(E)$ is increasing. 

Assuming that harvesting effort is at $E_{Y_m}$, the corresponding equilibrium values are  $\hat{x}(E_{Y_m}) = \displaystyle\frac{cr + be}{ac + bd} - \frac{cq_1 - bq_2}{ac + bd}E_{Y_m}$ and $\hat{y}(E_{Y_m}) = \displaystyle\frac{rd -ae}{ac + bd} - \frac{dq_1 + aq_2}{ac + bd}E_{Y_m}$. Hence, in order for MSY to exist, both $\hat{x}(E_{Y_m})>0$ and $\hat{y}(E_{Y_m}) > 0$. Substituting $E_{Y_m}$ gives us necessary conditions for the existence of MSY, summarized in the following Theorem.

\begin{theorem}
\label{theorem_both_harv}
In the system (\ref{general_Bionomic_system}), the following cases are possible:
\begin{itemize}
    \item if $cq_1 > bq_2$, then MSY may be obtained as a result of harvesting both prey and predator with some effort level such that $$E_{MSY} < \min\left(\dfrac{cr + be}{cq_1 - bq_2}, \dfrac{rd - ae}{dq_1 + aq_2}\right),$$
    \item if $cq_1 < bq_2$, then MSY may be obtained as a result of harvesting both prey and predator with some effort level such that $E_{MSY} < \dfrac{rd - ae}{dq_1 + aq_2}$.
\end{itemize}
\end{theorem}

\begin{figure}[!ht]
    \centering
    \includegraphics[scale=0.6]{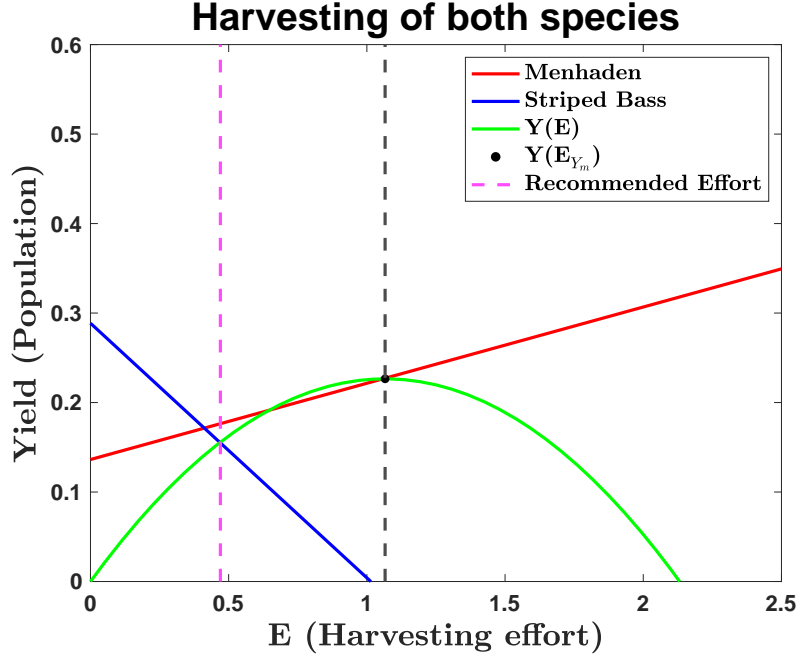}
    \caption{Yield function and the species equilibria curves for menhaden and bass when both prey and predator are harvested and parameters from Table \ref{ParmsTable} are used, with $q_1 = 0.5$ and $q_2 = 1$.}
    \label{Figure_Prey_Predator_Harvesting}
\end{figure}

Recall from Section \ref{3.4 case 3} Case 3, that for our menhaden-bass system based on parameters values, we have the second case.  Figure (\ref{Figure_Prey_Predator_Harvesting}) shows the yield and equilibria curves for the menhaden and bass using the parameters from Table \ref{ParmsTable}. Simultaneous prey and predator harvesting reduces the population of the predator while the population of the prey increases at equilibrium. As $\hat{y}(E_{Y_m})$ is negative, harvesting at $E_{Y_m}$ is not sustainable. Thus for our menhaden-bass system, according to Theorem \ref{theorem_both_harv}, $E_{MSY} <  \dfrac{rd - ae}{dq_1 + aq_2}\approx 1.02$. However, to ensure that the predator species are not endangered and driven to extinction, it is suggested for the maximum sustainable effort to be such that the population is decreased no more than 50\%;  hence, the suggested MSY in this case is $E_{MSY} = 0.5$, which is illustrated by the magenta line in Figure  (\ref{Figure_Prey_Predator_Harvesting}).

\subsection{Maximum Economic Yield}

Although the implementation of an MSY policy may theoretically be useful in sustaining fish populations while maximizing the yield, this policy has its limitations. A potential issue with harvesting at MSY is that due to inherent uncertainty in data and models, harvesting at the theoretical effort corresponding to MSY may result in overharvesting - leading to at risk or extinct species. This is where a different policy, maximum economic yield (MEY), can help minimize the risk. MEY is defined as the largest difference between total revenue earned from fishing and total costs of fishing. In other words, MEY is the largest net profit, or rent, that is earned from fishing. In order to understand why MEY may be a better option over MSY, it is important to analyze how they theoretically compare to one another.

\subsubsection*{Case 1. Harvesting the predator only}
MEY is defined as the largest difference between the total revenue and total costs. The revenue as a function of the fishing effort $E_2 = E$ is composed of the yield equation for harvesting the predator only, defined in (\ref{yield_predator_harvesting}), times $\hat{p}$, the price per unit of fish, and is given by the following equation
$$
    R(E) = Y(E) \hat{p} = E\left(\frac{rd-ae}{ac+bd} - \frac{aq_2}{ac+bd}E\right)\hat{p}.
$$
The cost function $ C(E) = \hat{c}E$ is
defined by the cost per unit of effort, $\hat{c}$, times harvesting effort $E$. We need to mention here that there may be some costs that are independent of effort, such as mortgage payments on a boat or rental of dock space for a boat, which are not considered in this equation. 
The rent function then is the difference between total revenue and total costs $Rent(E) = R(E) - C(E)$, which is also the objective in MEY policy. 
The rent function, when only the predator is harvested, is given by 
\begin{equation}\label{RentPred}
Rent(E) = \frac{\hat{p}(rd-ae)}{ac+bd}E - \frac{\hat{p}(aq)}{ac+bd}E^2 - \hat{c}E,
\end{equation}
which is a quadratic function of $E$. To maximize the rent function, we need to find the critical points of (\ref{RentPred}). For this purpose, let us find the derivative
$$
\frac{d}{dE}(Rent(E)) = \frac{\hat{p}(rd-ae)}{ac+bd} - \frac{2\hat{p}aq}{ac+bd}E - \hat{c} = 0,
$$
and solve for $E$. As the second derivative is always negative, this critical point gives us the harvesting effort that maximizes the rent 
\begin{equation}
    E_{MEY} = \frac{\hat{p}(rd-ae)-\hat{c}(ac+bd)}{2\hat{p}aq_2}.
\end{equation}
We still require that this effort is positive $E_{MEY} > 0$, which will be true as long as $\dfrac{rd-ae}{ac+bd} > \dfrac{\hat{c}}{\hat{p}}$.

Next, we want to compare $E_{MEY}$ with $E_{MSY}$ for the case when only the predator is harvested. According to Theorem \ref{theorem_pred_harv}, MSY exists and 
is given by $E_{MSY} = \dfrac{rd-ae}{2aq_2}$ if $rd > ae$.
Let us rewrite $E_{MEY}$ as follows 
\begin{equation}\label{comparison_1}
    E_{MEY} = \frac{rd-ae}{2aq_2} - \dfrac{\hat{c}(ac+bd)}{2\hat{p}aq_2} = E_{MSY} - \dfrac{\hat{c}(ac+bd)}{2\hat{p}aq_2}.
\end{equation}
From  (\ref{comparison_1}) we see that $E_{MEY} < E_{MSY}$, meaning that when harvesting the predator only, it takes less effort to achieve maximum economic yield than it does to achieve maximum sustainable yield. Note also that $E_{MEY}$ approaches $E_{MSY}$ as price $\hat{p}$ increases and as cost $\hat{c}$ decreases. Furthermore, MEY is still sustainable.

\subsubsection*{Case 2. Harvesting the prey only}
In the case when only the prey is harvested, using the yield function from (\ref{yield_prey_harvesting}), the rent function becomes
$$
Rent(E) = \frac{(cr+be)\hat{p}}{ac+bd}E - \frac{\hat{p}cq_1}{ac+bd}E^2 - \hat{c}E.
$$
To find $E$ that maximizes the rent, let us find the derivative 
$$
\frac{d}{dE}(Rent(E)) = \frac{(cr+be)\hat{p}}{ac+bd} - \frac{2\hat{p}cq_1}{ac+bd}E - \hat{c} = 0, 
$$
and solve the equation for $E$. As the second derivative is always negative, this means that the found critical point is the effort that maximizes the rent, or $E_{MEY}$, given by
\begin{equation}
    E_{MEY} = \frac{(cr+be)\hat{p}-\hat{c}(ac+bd)}{2\hat{p}cq_1}.
\end{equation}
Requiring positivity of $E_{MEY}$ gives the following condition on the system's parameters 
$\dfrac{cr+be}{ac+bd} > \dfrac{\hat{c}}{\hat{p}}$.
Similarly to the case of harvesting the predator only, we want to compare $E_{MEY}$ and $E_{MSY}$, from Theorem \ref{theorem_prey_harv} under the conditions of the theorem. By modifying $E_{MEY}$:
\begin{equation}\label{comparison_2}
    E_{MEY} = \frac{cr+be}{2cq_1} - \frac{\hat{c}(ac+bd)}{2\hat{p}cq_1} = E_{MSY} - \frac{\hat{c}(ac+bd)}{2\hat{p}cq_1}.
\end{equation}
As we can see from (\ref{comparison_2}), $E_{MEY} < E_{MSY}$ for harvesting the prey only, meaning that it takes less effort to achieve maximum economic yield than it does to achieve maximum sustainable yield. 

Here we must remember that the effort that maximizes the rent is less the effort that maximizes the yield ($E_{MEY} < E_{MSY} = E_{Y_m}$) only if the parameters of the system (\ref{general_Bionomic_system}) satisfy the relationship $\dfrac{rd - ae}{cr+be} > \dfrac{d}{2c}$. For the menhaden-bass system under consideration, this condition is not true. Furthermore, using the parameters' values from Table \ref{ParmsTable} and assuming $\hat{p} = 2\hat{c}$ for simplicity, we obtained that $E_{MEY}$ is negative and hence is not a valid harvesting effort. The optimal harvesting effort in the case of menhaden-bass system will be the recommended $E_{MSY} = \dfrac{rd-ae}{2dq_1} = 0.5$.

\subsubsection*{Case 3. Combined harvesting of both the prey and predator}
In the case of combined harvesting of both prey and predator with a harvesting effort $E_1=E_2=E$, the rent function becomes
$$
Rent(E) = \frac{\hat{p}(cr+be+rd-ae)}{ac+bd}E - \frac{\hat{p}(cq_1-bq_2+dq_1+aq_2)}{ac+bd}E^2 - \hat{c}E,
$$
with the yield function from (\ref{yield_combined_harvest}). Solving for critical points by setting the derivative with respect to $E$ to zero
$$
\frac{d}{dE}(Rent(E)) = \frac{\hat{p}(cr+be+rd-ae)}{ac+bd} - \frac{2\hat{p}(cq_1-bq_2+dq_1+aq_2)}{ac+bd}E - \hat{c} = 0
$$
gives us the MEY harvesting effort
\begin{equation}
    E_{MEY} = \frac{\hat{p}(cr+be+rd-ae)-\hat{c}(ac+bd)}{2\hat{p}(cq_1-bq_2+dq_1+aq_2)},
\end{equation}
if the obtained value is positive. In this case, both $E_{MSY}<E_{MEY}$ and $E_{MSY} > E_{MEY}$ are possible depending on the model's parameters. For the menhaden-bass system, using the values from Table \ref{ParmsTable}, we obtained a negative harvesting effort for $E_{MEY}$, which is not a valid option.

\subsection{Resilience Maximizing Yield}

Resilience maximizing yield (RMY) is a management approach that prioritizes resilience, a measure of how fast the system stabilizes back to biological equilibrium after some disturbance. Doing so allows fisheries to preserve the longevity and stability of the fish populations under consideration. Here, we analyze the effect of harvesting on the yield and resilience of the predator-prey system. The plots feature varying levels of the relevant catchability coefficient $q_i$ ($i = 1,2$), representative of how ``hard" it is to catch the fish, to study their effect on the system's resilience.

\subsubsection*{Case 1. Harvesting the predator only}

\begin{figure}[!ht]
    \centering
   \includegraphics[width=\textwidth]{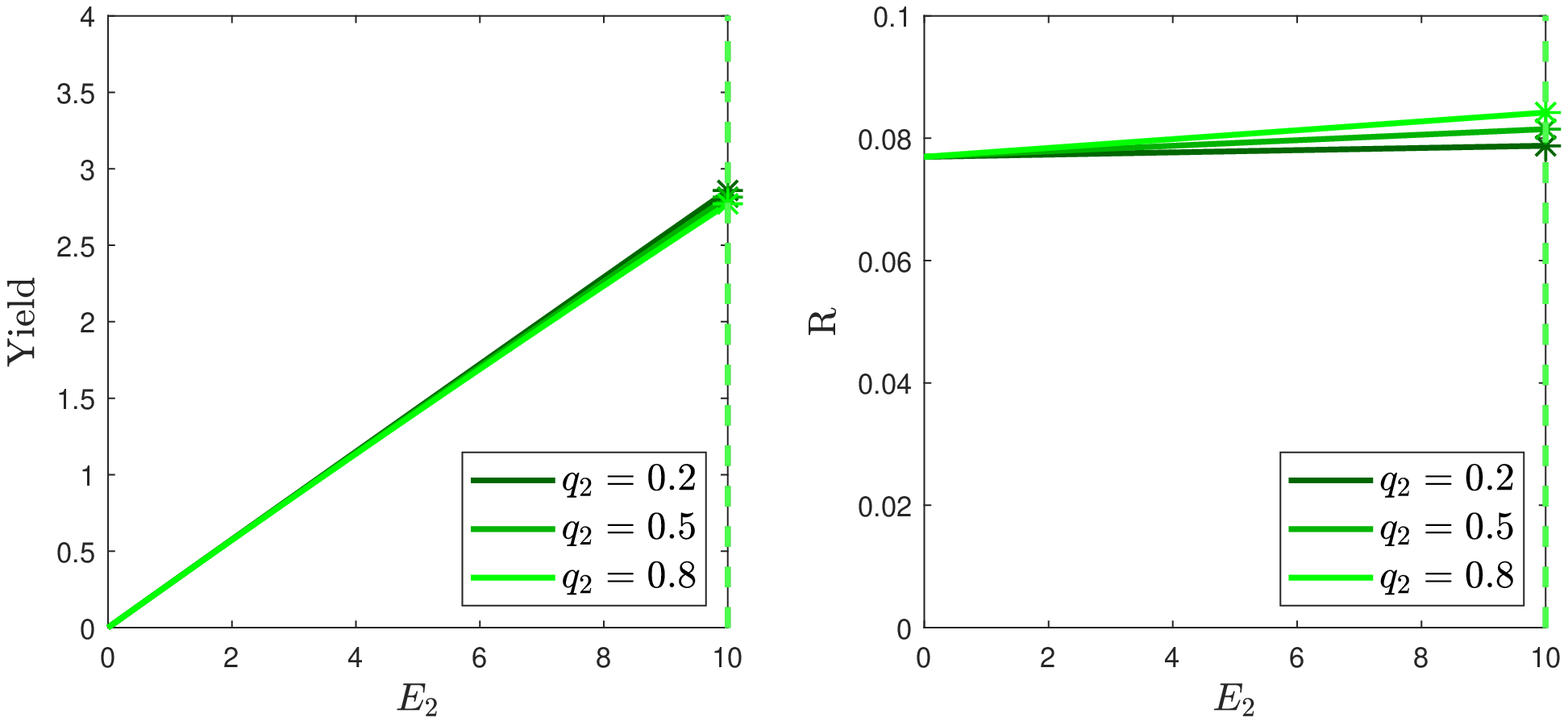}
    \caption{Yield curves (left) and resilience (right) for harvesting the predator only given by \ref{yield_predator_harvesting}, with varying values of $q_2$. Stars indicate the maximum of the resilience curves with their counterparts plotted on the corresponding yield curves.}
    \label{predator harvesting resilience and yield2}
\end{figure}

When only harvesting the striped bass, we want to see the effect of increasing the harvesting effort on the menhaden-bass system's resilience. As we set an upper bound for all model's parameters to 10, we also set a bound for the harvesting effort by considering the following range for $0 \le E_2 \le 10$. As the effort increases, the resilience curves increase linearly with the yield curves all being downward facing non-negative parabolas as illustrated in Figure \ref{predator harvesting resilience and yield2}. With our imposed bound on $E_2$, resilience can be maximized by harvesting at $E_{RMY} = 10$.

\subsubsection*{Case 2. Harvesting the prey only}

\begin{figure}[!ht]
    \centering
    \includegraphics[width=\textwidth]{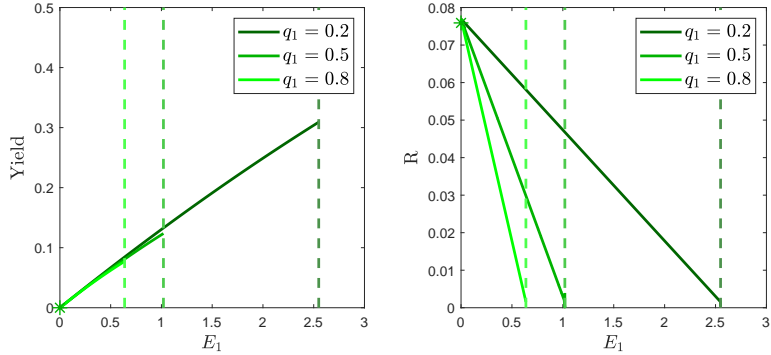}
    \caption{Yield curves (left) and resilience (right) for harvesting the prey only given by \ref{yield_prey_harvesting}, with varying values of $q_1$. Stars indicate the maximum of the resilience curves with their counterparts plotted on the corresponding yield curves.}
    \label{prey harvesting resilience and yield2}
\end{figure}

When only harvesting the menhaden, the resilience curves are strictly decreasing, as seen from Figure \ref{prey harvesting resilience and yield2}, with corresponding downward facing non-negative parabola yield curves. As the resilience curves are strictly decreasing, resilience is maximized at $E_1 = 0$, showing that any additional effort on harvesting the prey is at the expense of the system's resilience. However, additional yield and profit can be obtained at the sacrifice of the system's biological stability with $E_1 > 0$.

\subsubsection*{Case 3. Combined harvesting of both the prey and the predator}

\begin{figure}[!ht]
    \centering
    \includegraphics[width=\textwidth]{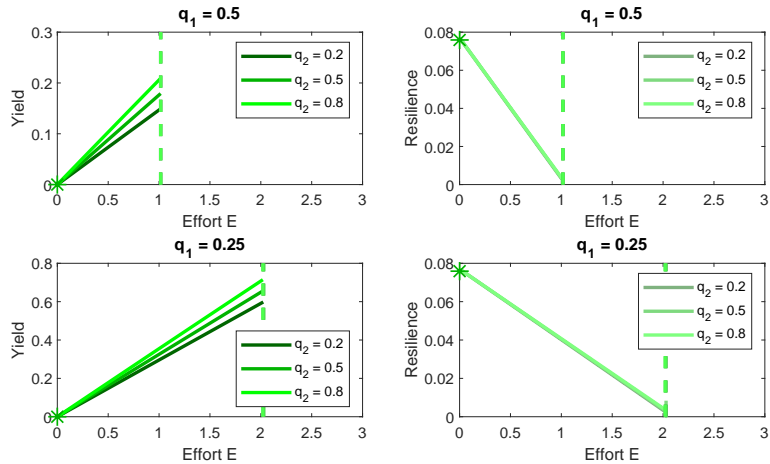}
   \caption{Yield curves (left) and resilience (right) for harvesting both species given by \ref{yield_combined_harvest}, with varying values of $q_2$ when $q_1 = 0.5, 0.25$. Stars indicate the maximum of the resilience curves with their counterparts plotted on the corresponding yield curves.}
    \label{Both harvesting change q2 for two q1}
\end{figure}

\begin{figure}[!ht]
    \centering
    \includegraphics[width=\textwidth]{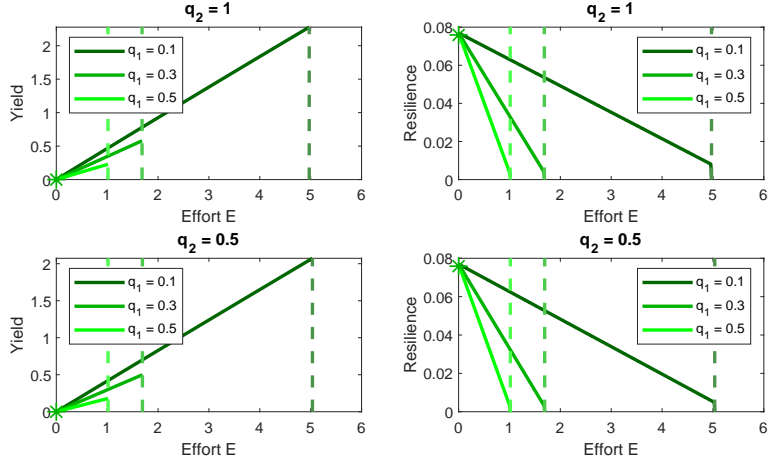}
    \caption{Yield curves (left) and resilience (right) for harvesting both species given by \ref{yield_combined_harvest}, with varying values of $q_1$ when $q_2 = 1, 0.5$. Stars indicate the maximum of the resilience curves with their counterparts plotted on the corresponding yield curves.}
    \label{Both harvesting change q1 for two q2}
\end{figure}

When harvesting both menhaden and bass at the same effort level, we see that the general behavior of the resilience and yield curves are uniform with $q_i$. In Figure \ref{Both harvesting change q2 for two q1}, all of the resilience curves are strictly decreasing. Hence, the effort must be zero $E_{RMY} = 0$ to maximize the resilience. When both $q_1 = 0.5, 0.25$, further effort towards harvesting both species can be applied for additional yield at the expense of the resilience of the system. In Figure \ref{Both harvesting change q1 for two q2}, we similarly have that the resilience curves are strictly decreasing and $E_{RMY} = 0$. When both $q_2 = 1, 0.5$, additional simultaneous harvesting effort results in increased yield at the cost of resilience.

Our analysis also shows that varying the $q_i$ can have significant economical implications when considering RMY. Vertical dashed lines indicate when the resilience of the system is equal to 0, meaning that the system can no longer recover as one of the species has gone extinct. Stars indicate the maximum of the resilience curves at effort levels $E_{RMY}$, with their counterparts plotted on the corresponding yield curves, with coordinates ($E_{RMY}$, $Y(E_{RMY})$). Note that the resilience curves in all cases have initial value of approximately the original resilience of the system without harvesting from \ref{Tau}, for which the return time to equilibrium is about 13 years.

\section{Sensitivity analysis}\label{sec5}
Another important consideration is if the solutions to the system of differential equations are stable with respect to small perturbations of the parameters of the system. To answer this question, we perform sensitivity analysis; this gives us an idea of how small changes in the system's parameters affect the behavior of the solutions in time. We use the direct differential method, which is a local method for sensitivity analysis.

First, assume that the mathematical model, consisting of the first two differential equations in (\ref{general_Bionomic_system}), has the form
\begin{equation}
    \frac{du}{dt} = f(u, p),
\end{equation}
where $u=u(t,p) = [x(t,p) \quad y(t,p)]$ is the two component vector solution of the system, $p =(p_1,\dots, p_n)$ is the vector of all parameters, and the $f$ is the two component vector of the right-hand sides. We are looking to find the rate of change of $u$ with respect to $p_j$, i.e. $\dfrac{\partial u}{\partial p_j}$. We can find the time evolution of this quantity by using the given differential equation as follows
\begin{equation}\label{sensitivity_equation}
\frac{d}{dt}\left(\frac{\partial u_{i}}{\partial p_{j}}\right) =\frac{\partial f_{i}}{\partial u_{i}}
\frac{\partial u_{i}}{\partial p_{j}} + \frac{\partial f_{i}}{\partial p_{j}} = J \cdot S_{j} +F_{j},
\end{equation}
where matrix $J$, the Jacobian, is given by
$$
J = \left(
\def\arraystretch{1.4}
\begin{array}{cccccc}
\frac{\partial f_{1}}{\partial u_{1}} & \frac{\partial f_{1}}{\partial u_{2}} & ... & \frac{\partial f_{1}}{\partial u_{k}} \\
\frac{\partial f_{2}}{\partial u_{1}} & \frac{\partial f_{2}}{\partial u_{2}} & ... & \frac{\partial f_{2}}{\partial u_{k}} \\
... & ... & ... & ...\\
\frac{\partial f_{k}}{\partial u_{1}} & \frac{\partial f_{k}}{\partial u_{2}} & ... & \frac{\partial f_{k}}{\partial u_{k}} \\
\end{array}
\right),
$$
vector $S_{j}$, the sensitivity vector for parameter $p_{j}$, and vector $F_{j}$, the derivative of the right-hand side of the system with respect to the parameter $p_{j}$, are given by the vectors:
$$
S_{j} = \left(\frac{\partial y_{i}}{\partial p_{j}}\right) = 
\left(
\def\arraystretch{1.4}
\begin{array}{c}
\frac{\partial y_{1}}{\partial p_{j}}\\
\frac{\partial y_{2}}{\partial p_{j}}\\
...\\
\frac{\partial y_{k}}{\partial p_{j}}
\end{array}
\right),
F_{j} =\left(
\def\arraystretch{1.4}
\begin{array}{c}
\frac{\partial f_{1}}{\partial p_{j}}  \\
\frac{\partial f_{2}}{\partial p_{j}}\\
...\\
\frac{\partial f_{k}}{\partial p_{j}}
\end{array}
\right).
$$
The Jacobian for our system from (\ref{general_Bionomic_system}) is given by the matrix
$$
J(x,y) = 
\left(
\begin{array}{ccc}
r - 2ax -by - q_1E_1 & -bx \\
dy & -e - 2cy + dx - q_2 E_2 
\end{array}
\right).
$$
As the system (\ref{general_Bionomic_system}) has 8 different parameters, let us arrange all parameters as a vector $p = \left( 
r \:\:\: a \:\:\: b \:\:\: E_1 \:\:\: 
e\:\:\: c \:\:\: d\:\:\: E_2 
\right).$
Then the full sensitivity matrix $S$ is given by
\begin{equation}\label{SensitivityMatrix}
S = \left(
\def\arraystretch{1.4}
\begin{array}{cccccccc}
\frac{\partial x}{\partial r} & \frac{\partial x}{\partial a} & 
\frac{\partial x}{\partial b} & \frac{\partial x}{\partial E_1} & 
\frac{\partial x}{\partial e} & \frac{\partial x}{\partial c} & 
\frac{\partial x}{\partial d} & \frac{\partial x}{\partial E_2}\\
\frac{\partial y}{\partial r} & \frac{\partial y}{\partial a} & 
\frac{\partial y}{\partial b} & \frac{\partial y}{\partial E_1} & 
\frac{\partial y}{\partial e} &\frac{\partial y}{\partial c} & 
\frac{\partial y}{\partial d} & \frac{\partial y}{\partial E_2} \\
\end{array}
\right).
\end{equation}
And the full derivative of the right-hand side with respect to the parameters is given by the matrix
\begin{equation}\label{fullFDD}
F = 
\left(
\begin{array}{cccccccccccccc}
x & -x^2 & -xy & -x & 0 & 0 & 0 & 0\\
0 & 0 & 0 & 0 & -y & -y^2 & xy & -y \\
\end{array}
\right).
\end{equation}
Hence, for example, the system of differential equations that investigates the rate of change of the dependent variables $x$ and $y$ with respect to the parameter $r$ is given by four differential equations:
$$
\left\{
\def\arraystretch{2}
\begin{array}{lll}
\dfrac{dx}{dt} & = & x(r-ax-by)-q_{1}E_{1}x  \\
\dfrac{dy}{dt} & = & y(-e-cy+dx)-q_{2}E_{2}y\\
\dfrac{ds_{11}}{dt} & =& (r -2ax - by - E_1)s_{11}  - bxs_{21} + x\\
\dfrac{ds_{21}}{dt} &=& dy s_{11} - (e + 2cy - dx -q_2E_2)s_{21}\\
\end{array}
\right.
$$
Observe that the first two differential equations are exactly the differential equations in the system (\ref{general_Bionomic_system}) while the third and fourth equations come from (\ref{sensitivity_equation}) by using the Jacobian and the first column of (\ref{fullFDD}). 
Since different parameters can have different units and thus can have different orders of magnitude, we calculate the relative sensitivity matrix $\bar{S}_{ij} = \dfrac{\partial u_i}{\partial p_j} \dfrac{p_j}{u_i}$. Then the sensitivity of each parameter $p_j$ is given by the sensitivity index defined as the magnitude of the corresponding norm of the $i,j$-th element in $\bar{S}$:
$$
\left\Vert\bar{S}_{ij}\right\Vert_2 = \sqrt{\sum_{k=1}^N \left(\dfrac{\partial u_i}{\partial p_j}(t_k) \dfrac{p_j}{u_i(t_k)}\right)^2}.
$$
For the parameter set given in Table \ref{ParmsTable}, we computed
the relative sensitivity vector for each of the steady-states of the variables $x(t)$ and $y(t)$ (for $2050<t<2100$) which are given in descending order below.

\begin{center}
\begin{tabular}{ c c}  
{Menhaden's Norms} & {Bass' Norms} \\
{\tabulinesep=1mm
   \begin{tabu} {|c c|}
       \hline
        Parameter p&$\left\Vert\frac{dx}{dp}\right\Vert_{2}$\\
        \hline
        d&0.9985\\ 
        \hline
        e&0.8886\\ 
        \hline
        c&0.1100\\ 
        \hline
        b&0.1098\\ 
        \hline
        r&0.1079 \\
        \hline
        a&0.0008\\
        \hline
        $E_2$&$0.7328 E_2 q_2$\\
        \hline
        $E_1$&$0.2103 E_1 q_1$\\
        \hline
    \end{tabu}} & 
    {\tabulinesep=1mm
    \begin{tabu} {|c c|}
        \hline
        Parameter p&$\left\Vert\frac{dy}{dp}\right\Vert_{2}$\\
        \hline
        r&1.0080\\ 
        \hline
        b&1.0004\\ 
        \hline
        e&0.0133\\ 
        \hline
        d&0.0093\\
        \hline
        a&0.0069\\ 
        \hline
        c&0.0029\\
        \hline
        $E_1$&$1.9647 E_1 q_1$\\
        \hline
        $E_2$&$0.0110 E_2 q_2$\\
        \hline
   \end{tabu}}\\
\end{tabular}
\end{center}

\begin{figure}[!ht]
    \makebox[\textwidth][c]{\includegraphics[scale=0.64]{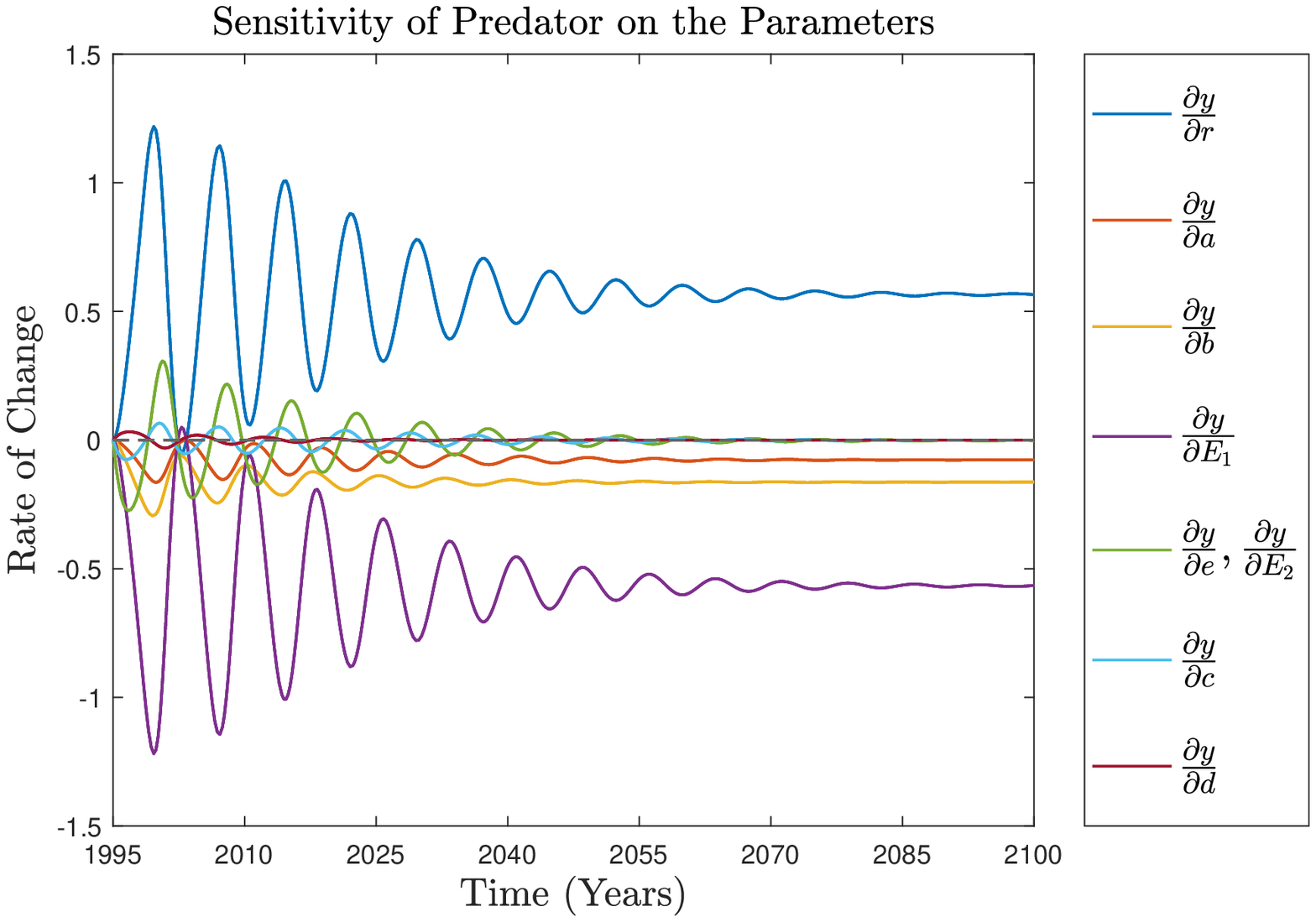}} 
    \makebox[\textwidth][c]{\includegraphics[scale=0.64]{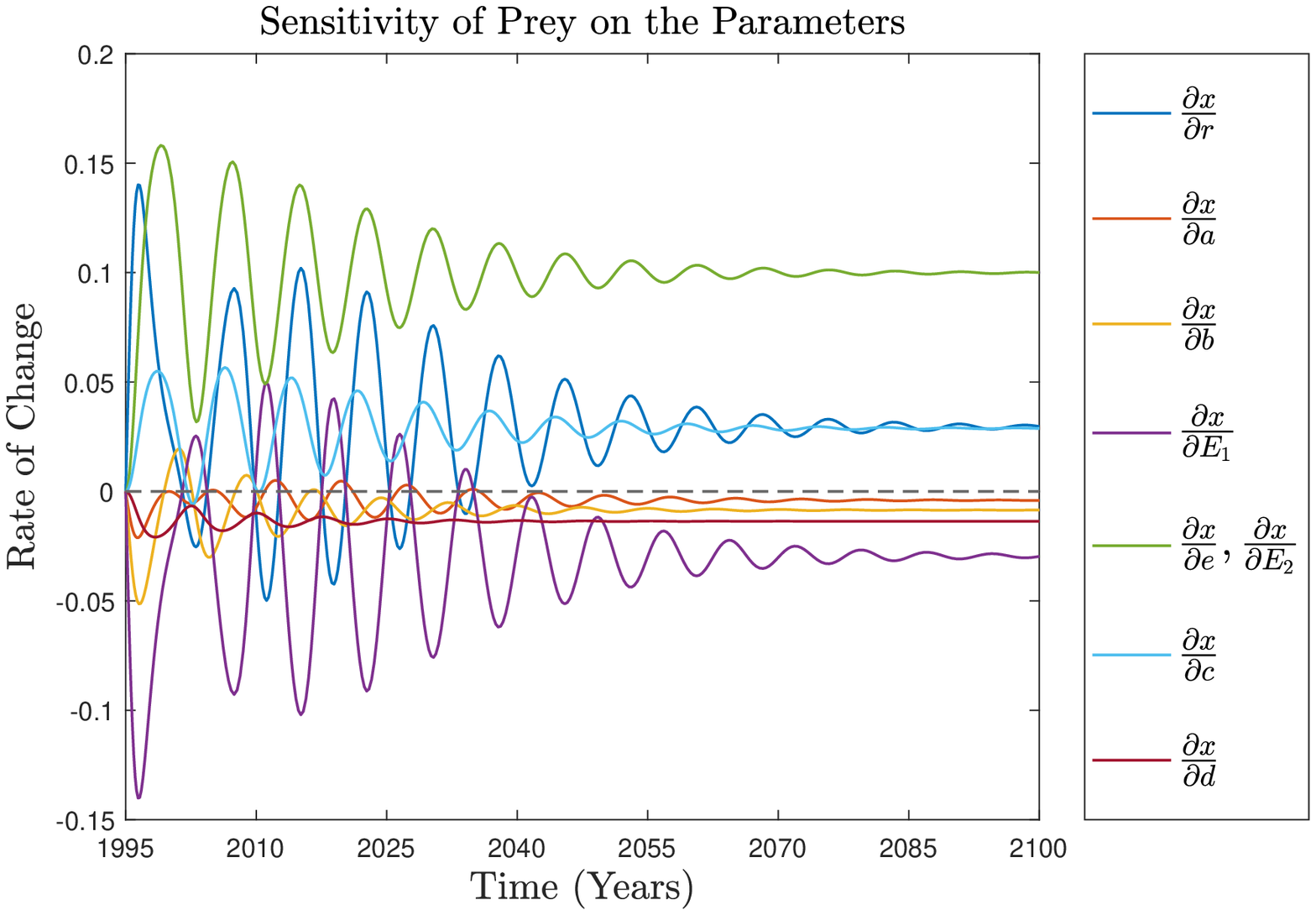}}
    \caption{The rate of change of each species' population with respect to the system parameters over time.}
    \label{sensitivity_rate_of_change}
\end{figure}

\begin{figure}[!ht]
    \makebox[\textwidth][c]{\includegraphics[scale=0.82]{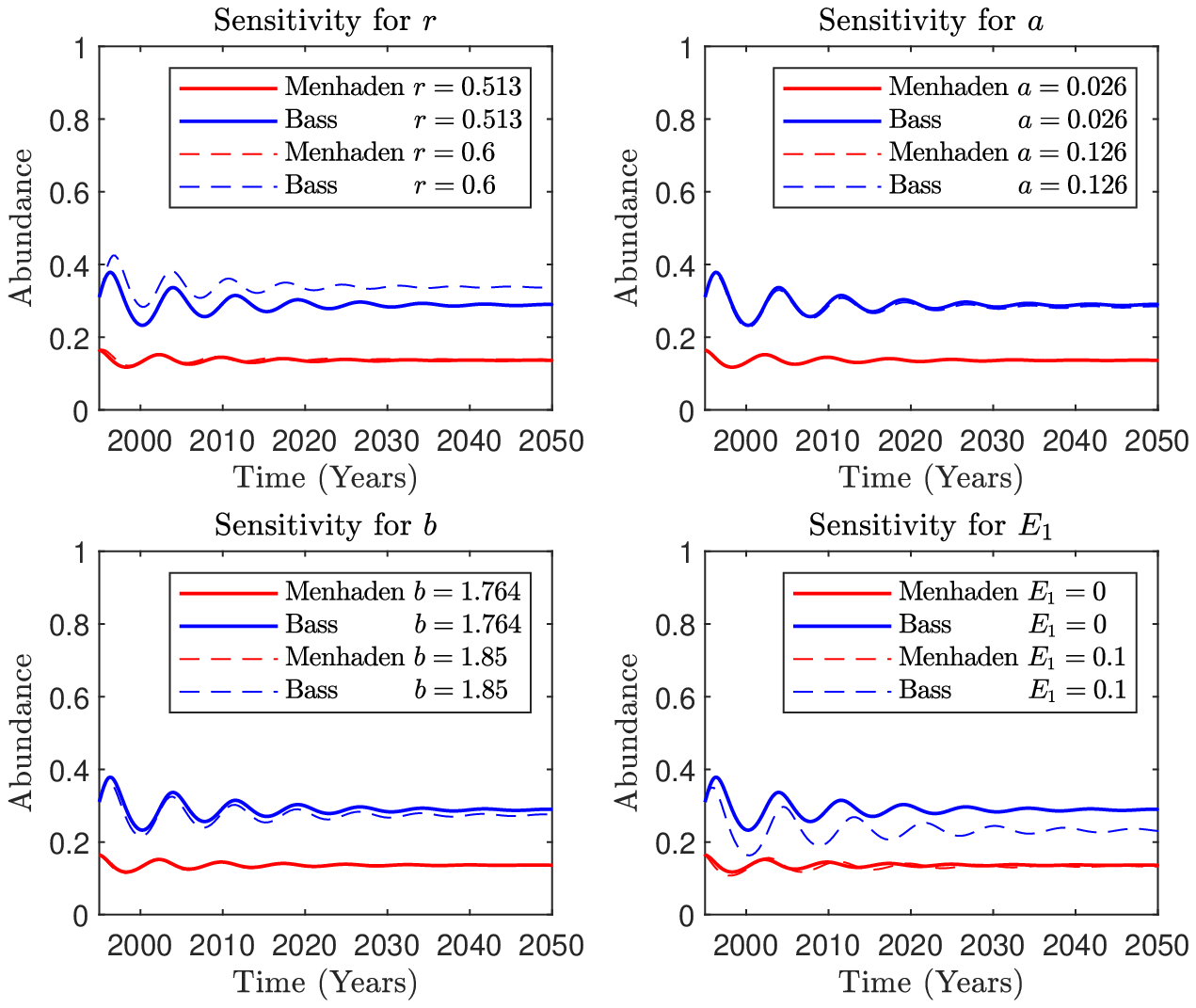}}
    \label{menhaden_sens}
    \vspace{-0.5cm}
    \makebox[\textwidth][c]{\includegraphics[scale=0.82]{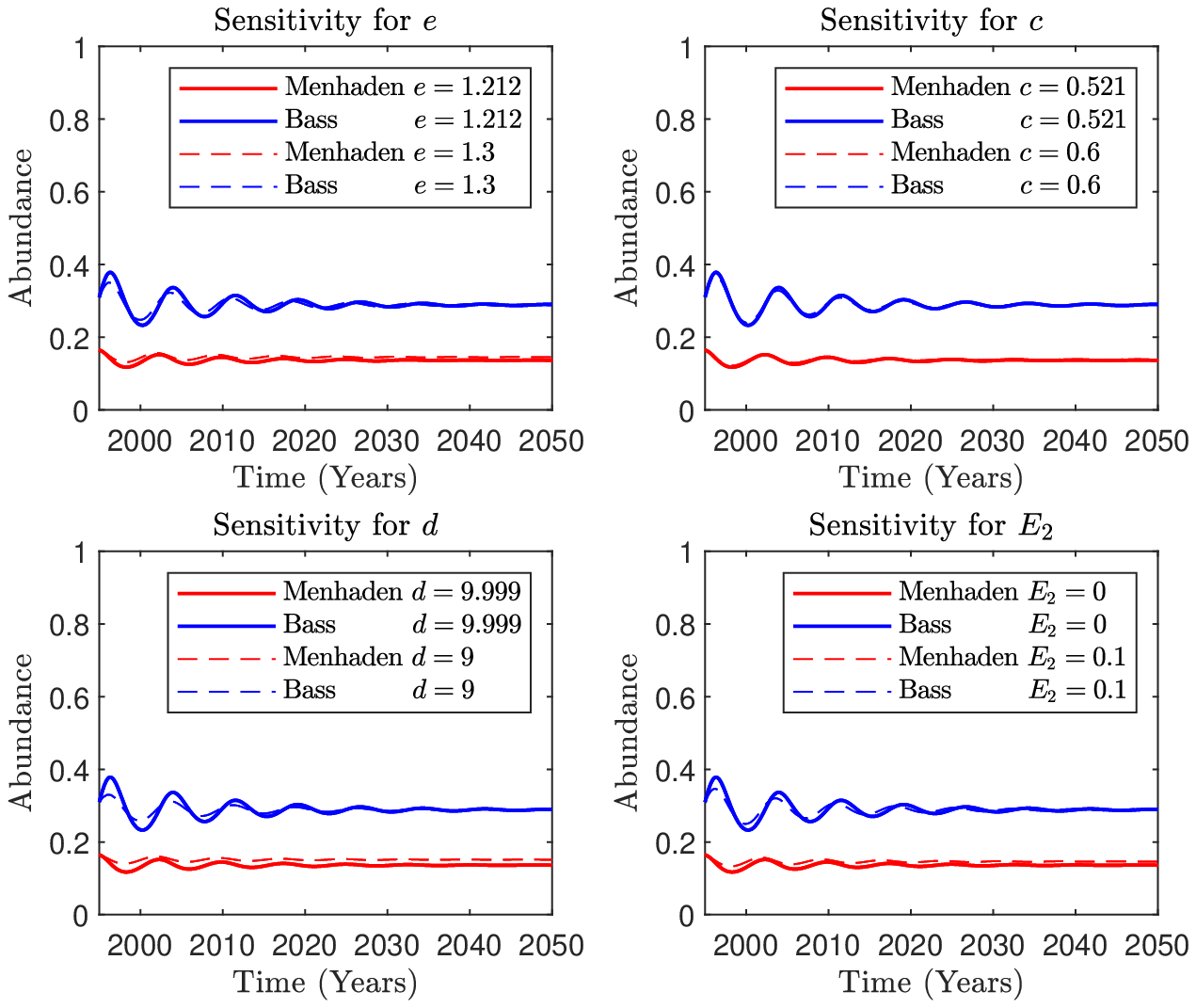}}
    \caption{State variables, menhaden in red and bass in blue, time evolution for small perturbations in the model parameters values from Table \ref{ParmsTable}. }
    \label{sensitivity_state_variables}
\end{figure}

Figure \ref{sensitivity_rate_of_change} illustrates the sensitivity of striped bass (the top graph) and Atlantic menhaden (bottom graph) with respect to each of the eight parameters by showing the rate of change of each variable with respect to the parameter over time. From the sensitivity tables and Figure \ref{sensitivity_rate_of_change}, we see that the striped bass is most sensitive to perturbations in the intrinsic growth rate of the Atlantic menhaden $r$. As illustrated in Figure \ref{sensitivity_state_variables}, a small increase in the growth rate of Atlantic menhaden increases the steady-state of bass significantly from about 0.29 to 0.34. From the sensitivity table for Atlantic menhaden, we see that it is the most sensitive to changes in the parameter $d$, the effect of prey consumption on bass. A marginal decrease of $d$ increases the steady-state of Atlantic menhaden but leaves the steady-state of the striped bass' population relatively unaffected. The next most sensitive parameter for menhaden is the natural mortality rate of striped bass $e$. Similarly, a marginal increase of $e$, as shown in Figure  \ref{sensitivity_state_variables}, results in a higher steady-state for Atlantic menhaden but does not significantly change the steady-state for the striped bass.

Other important parameters from the sensitivity tables are the harvesting coefficients $E_1$ and $E_2$. The predator (striped bass) is more sensitive to the prey's harvesting effort $E_1$, while the prey (Atlantic menhaden) is more sensitive to the predator's harvesting effort $E_2$. As illustrated in Figure \ref{sensitivity_state_variables}, a marginal increase of ${E_2}$ results in an increase of the prey's steady-state and has little effect on the predator's steady-state -- meaning that the predator's population is more sensitive to changes in the harvesting of the predator. We can also see in Figure \ref{sensitivity_state_variables} that a marginal increase of ${E_1}$ results in a decrease in the steady-state of the bass' population with little effect on the Atlantic menhaden.

\section{Conclusions}\label{sec6}
The Atlantic menhaden and striped bass are two of the most important fish in the Chesapeake Bay. Yet, both have experienced significant population decline  over the last few decades due to multiple factors, one of which is overharvesting. There is still uncertainty among policymakers about the management approach which would preserve both species. In this study, we applied mathematical modeling to investigate the current predator-prey dynamics between the species, and particularly the long-term effects of three harvesting policies: maximum sustainable yield (MSY), maximum economic yield (MEY), and resilience maximizing yield (RMY) within a two-species dynamical model. Specifically, we examine how each harvesting policy affects the long term population dynamics and explore the trade-off between maximizing economic yield and maintaining long-term sustainability within each policy.

We used the Lotka-Volterra predator-prey model to analyze the population dynamics of Atlantic menhaden and striped bass as well as the possibility of balancing economic gains with ecosystem conservation. We began by fitting the generalized Lotka-Volterra model to the actual time series data of the Atlantic menhaden and striped bass' adjusted abundance data to find the model's parameter values and calculate the resilience of the system. Next, we included a harvesting effort component into the model and found the bioeconomic equilibria as well as derived conditions for their stability. We have proven the positivity and boundedness of the solutions within an invariant region. This shows that the model is well posed, and if the initial conditions of the system are within this region, then all solutions will remain there. We studied the effects of maximum sustainable yield (MSY), maximum economic yield (MEY), and resilience maximizing yield (RMY) policies on the sustainability of the fish populations.

In our study, we have shown that maximum sustainable yield, defined as harvesting at an effort that maximizes the yield sustainably, could be achieved for both species in our predator-prey system when only the predator is harvested. In the case when only the prey is harvested, maximum sustainable effort is much lower than the effort at which yield is maximized. Similarly, when both species are harvested with the same effort, we found sufficient conditions under which the maximum sustainable effort exists and allows for sustainability of the two populations, but this effort is lower than the effort needed to maximize the yield.

Additionally, we have shown that when harvesting the predator only, the harvesting effort that maximizes the yield sustainably, $E_{MSY}$, is larger than the harvesting effort that maximizes the profit, $E_{MEY}$. Hence, it is not economically justified to harvest above $E_{MEY}$ as it does not increase the profit but adds additional costs and further decreases the fish populations. When harvesting the prey only or both prey and predator, the equilibrium population of the striped bass decreases at much higher rate compared to Atlantic menhaden with respect to $E$. In these cases, we have shown that both are possible: $E_{MSY}$ could be more or less than  $E_{MEY}$ based on the model's parameters. For the menhaden-bass system under consideration, with our parameters and assumed catchability, price, and cost coefficients, $E_{MEY}$ was negative; hence, MEY is not a possible management strategy.

When harvesting the predator only, resilience is initially increasing with $E$, showing that additional effort on harvesting the predator can increase the system’s resilience while also providing yield. When harvesting the prey only, resilience is maximized when the effort is strictly zero, showing that additional effort on harvesting the prey harms the system’s resilience. In this case, additional yield and profit can be obtained by sacrificing of the system’s biological resilience. When harvesting both the predator and prey simultaneously, the resilience curves and corresponding analysis become dependent on the catchability parameters. In either case, the resilience of the system declines as the effort increases, resulting in a similar yield-resilience trade-off.

Finally, we performed sensitivity analysis of the mathematical model with respect to small perturbations in its parameters. Sensitivity analysis revealed that the striped bass is most sensitive to the intrinsic growth rate of the Atlantic menhaden, $r$, and even a marginal increase of the growth rate of the prey would benefit the predator and increase the predator's steady-state significantly in the long term. The sensitivity analysis also has shown that both the Atlantic menhaden and striped bass are very sensitive to the coefficient $d$, the effect of prey consumption. A marginal decrease of $d$ increases the steady-state of Atlantic menhaden but leaves the steady-state of the bass’ population relatively unaffected. Our sensitivity analysis also shows that harvesting the prey affects the predator much more negatively than the prey, while harvesting the predator increases the prey's population steady-state while leaving the predator's largely unaffected. These sensitivity results shed light on the effects of the system's parameters on the long-term population dynamics of Atlantic menhaden and striped bass, and in conjunction with the results on the different harvesting policies, may be very beneficial for policy-makers when deciding on future fishing quotes for these species.

\section*{Acknowledgments}
We wish to acknowledge the support of the NSF-CURM grant DMS-1722563. The authors would like to thank Dr. M. Hallare for the fruitful discussions throughout the CURM experience.

\end{document}